\documentclass[twocolumn,traditabstract]{aa}
\usepackage{fixltx2e}
\usepackage[english]{babel}
\usepackage{graphicx,amsmath}
\usepackage{epsf,color}
\usepackage[mathscr]{eucal}
\usepackage{amsmath}
\usepackage{amssymb,amsfonts}
\usepackage{natbib}
\usepackage{txfonts}
\usepackage{dsfont}
\definecolor{Mygreen}{rgb}{0.00, 0.72, 0.0}
\definecolor{Mypink}{rgb}{1.0, 0.0, 0.5}
\usepackage[breaklinks, citecolor=blue, linkcolor=Mygreen, urlcolor=Mypink, colorlinks=true, debug, baseurl=' ']{hyperref}
\usepackage{etoolbox}
\makeatletter
\patchcmd\@combinedblfloats{\box\@outputbox}{\unvbox\@outputbox}{}{%
   \errmessage{\noexpand\@combinedblfloats could not be patched}%
}%
 \makeatother

\def\simlt{\lower.5ex\hbox{$\; \buildrel < \over \sim \;$}}
\def\simgt{\lower.5ex\hbox{$\; \buildrel > \over \sim \;$}}
\def\simgt{\lower.5ex\hbox{$\; \buildrel $\textgreater$ \over \sim \;$}}
\def\NIKA{\textit{NIKA}}
\def\NIKAd{\textit{NIKA2}}
\def\Archeops{\textit{Archeops}}
\def\Planck{\textit{Planck}}
\def\WMAP{\textit{WMAP}}

\bibpunct{(}{)}{;}{a}{}{,}
\begin{document}

\title{NIKA 150 GHz polarization observations of the Crab nebula and its spectral energy distribution}
\author{A.~Ritacco \inst{\ref{LPSC}},\inst{\ref{IRAME}}\thanks{Corresponding author: Alessia Ritacco, \url{ritaccoa@iram.es}}
\and  J.F.~Mac\'ias-P\'erez \inst{\ref{LPSC}}
\and  N.~Ponthieu \inst{\ref{IPAG}}
\and  R.~Adam \inst{\ref{LPSC},\ref{OCA},\ref{CEFCA}}
\and  P.~Ade \inst{\ref{Cardiff}}
\and  P.~Andr\'e \inst{\ref{CEA}}
\and  J.~Aumont \inst{\ref{IRAP2}}
\and  A.~Beelen \inst{\ref{IAS}}
\and  A.~Beno\^it \inst{\ref{Neel}}
\and  A.~Bideaud \inst{\ref{Neel}}
\and  N.~Billot \inst{\ref{IRAME}}
\and  O.~Bourrion \inst{\ref{LPSC}}
\and  A.~Bracco \inst{\ref{CEA},\ref{Nordita}}
\and  M.~Calvo \inst{\ref{Neel}}
\and  A.~Catalano \inst{\ref{LPSC}}
\and  G.~Coiffard \inst{\ref{IRAMF}}
\and  B.~Comis \inst{\ref{LPSC}}
\and  A.~D'Addabbo \inst{\ref{Neel},\ref{Roma}}
\and  M.~De Petris \inst{\ref{Roma}}
\and  F.-X.~D\'esert \inst{\ref{IPAG}}
\and  S.~Doyle \inst{\ref{Cardiff}}
\and  J.~Goupy \inst{\ref{Neel}}
\and  C.~Kramer \inst{\ref{IRAME}}
\and  G.~Lagache \inst{\ref{LAM}}
\and  S.~Leclercq \inst{\ref{IRAMF}}
\and  J.-F.~Lestrade \inst{\ref{LERMA}}
\and  P.~Mauskopf \inst{\ref{Cardiff},\ref{Arizona}}
\and  F.~Mayet \inst{\ref{LPSC}}
\and  A.~Maury \inst{\ref{CEA}}
\and  A.~Monfardini \inst{\ref{Neel}}
\and  F.~Pajot \inst{\ref{IRAP2}}
\and  E.~Pascale \inst{\ref{Cardiff}}
\and  L.~Perotto \inst{\ref{LPSC}}
\and  G.~Pisano \inst{\ref{Cardiff}}
\and  M.~Rebolo-Iglesias\inst{\ref{LPSC}}
\and  V.~Rev\'eret \inst{\ref{CEA}}
\and  L.~Rodriguez \inst{\ref{CEA}}
\and  C.~Romero \inst{\ref{IRAMF}}
\and  H.~Roussel \inst{\ref{IAP}}
\and  F.~Ruppin \inst{\ref{LPSC}}
\and  K.~Schuster \inst{\ref{IRAMF}}
\and  A.~Sievers \inst{\ref{IRAME}}
\and  G.~Siringo \inst{\ref{ALMA}}
\and  C.~Thum \inst{\ref{IRAME}}
\and  S.~Triqueneaux \inst{\ref{Neel}}
\and  C.~Tucker \inst{\ref{Cardiff}}
\and  H.~Wiesemeyer \inst{\ref{MPBonn}}
\and  R.~Zylka \inst{\ref{IRAMF}}}

\institute{
Laboratoire de Physique Subatomique et de Cosmologie, Universit\'e Grenoble Alpes, CNRS/IN2P3, 53, avenue des Martyrs, Grenoble, France
  \label{LPSC}
  \and
  Laboratoire Lagrange, Universit\'e C\^ote d'Azur, Observatoire de la C\^ote d'Azur, CNRS, Blvd de l'Observatoire, CS 34229, 06304 Nice cedex 4, France
  \label{OCA}
  \and
Institut de RadioAstronomie Millim\'etrique (IRAM), Grenoble, France
  \label{IRAMF}
\and
Laboratoire AIM, CEA/IRFU, CNRS/INSU, Universit\'e Paris Diderot, CEA-Saclay, 91191 Gif-Sur-Yvette, France 
  \label{CEA}
\and
Astronomy Instrumentation Group, University of Cardiff, UK
  \label{Cardiff}
\and
Institut d'Astrophysique Spatiale (IAS), CNRS and Universit\'e Paris Sud, Orsay, France
  \label{IAS}
\and
Institut N\'eel, CNRS and Universit\'e Grenoble Alpes, France
  \label{Neel}
\and
Institut de RadioAstronomie Millim\'etrique (IRAM), Granada, Spain
  \label{IRAME}
\and
Dipartimento di Fisica, Sapienza Universit\`a di Roma, Piazzale Aldo Moro 5, I-00185 Roma, Italy
  \label{Roma}
\and
Univ. Grenoble Alpes, CNRS, IPAG, 38000 Grenoble, France 
  \label{IPAG}
    \and
Aix Marseille Universit\'e, CNRS, LAM (Laboratoire d'Astrophysique de Marseille) UMR 7326, 13388, Marseille, France
  \label{LAM}
\and
School of Earth and Space Exploration and Department of Physics, Arizona State University, Tempe, AZ 85287
  \label{Arizona}
\and
Universit\'e de Toulouse, UPS-OMP, Institut de Recherche en Astrophysique et Plan\'etologie (IRAP), Toulouse, France
  \label{IRAP}
\and
CNRS, IRAP, 9 Av. colonel Roche, BP 44346, F-31028 Toulouse cedex 4, France 
  \label{IRAP2}
\and
University College London, Department of Physics and Astronomy, Gower Street, London WC1E 6BT, UK
  \label{UCL}
\and  
Institut d'Astrophysique de Paris, Sorbonne Universit\'es,
  UPMC Univ. Paris 06, CNRS UMR 7095, 75014 Paris, France
\label{IAP}
\and
LERMA, CNRS, Observatoire de Paris, 61 avenue de l'Observatoire, Paris, France
  \label{LERMA}
  \and
  Centro de Estudios de F\'isica del Cosmos de Arag\'on (CEFCA), Plaza San Juan, 1, planta 2, E-44001, Teruel, Spain
  \label{CEFCA}
  \and
  Joint ALMA Observatory \& European Southern Observatory, Alonso de C\'ordova 3107, Vitacura, Santiago, Chile
\label{ALMA}
\and
Max Planck Institute for Radio Astronomy, 53111 Bonn, Germany
\label{MPBonn}
\and
Nordita, KTH Royal Institute of Technology and Stockholm University, Roslagstullsbacken 23, 10691 Stockholm, Sweden
\label{Nordita}}
 
\date{Received \today \ / Accepted --}
        
\abstract{
  The \object{Crab} nebula is a supernova remnant exhibiting a highly polarized synchrotron
  radiation at radio and millimetre wavelengths.
  It is the brightest source in the microwave sky with an extension
  of 7 by 5 arcminutes, and is commonly used as a standard candle for any experiment
  which aims to measure the polarization of the sky.
  Though its spectral energy distribution has been well characterized in total intensity,
  polarization data are still lacking at millimetre wavelengths.
  We report in this paper high resolution observations
  (18$^{\prime\prime}$ FWHM) of the \object{Crab} nebula in total intensity and
  linear polarization at 150 GHz with the \NIKA\ camera. \NIKA, operated at the IRAM
  30 m telescope from 2012 to 2015, is a camera made of Lumped Element Kinetic
  Inductance Detectors (LEKIDs) observing the sky at 150 and 260 GHz. From
  these observations we are able to reconstruct the spatial distribution of the
  polarization degree and angle of the \object{Crab} nebula, which is found to be compatible
  with previous observations at lower and higher frequencies. Averaging across
  the source and using other existing data sets we find that the \object{Crab} nebula
  polarization angle is consistent with being constant over a wide range of frequencies with a
  value of $-87.7^{\circ} \pm 0.3$ in Galactic coordinates.
  We also present the first estimation of the \object{Crab} nebula spectral energy distribution polarized flux
  in a wide frequency range: 30--353 GHz. Assuming a single power law emission model we find that the
  polarization spectral index $\beta_{pol}$= -- 0.347 $\pm$ 0.026 is compatible with the intensity
  spectral index $\beta$= -- 0.323 $\pm$ 0.001.}
\titlerunning{NIKA polarization observations of the
  Crab nebula.}  \authorrunning{A. Ritacco, J. F. Mac\'ias P\'erez , N. Ponthieu
  et al.}  \keywords{Techniques: Crab nebula -- Tau A -- polarization -- KIDs -- individual: NIKA }
\maketitle

\begin{figure*}[h!]
  \centering
          { \includegraphics[width=0.32\linewidth,keepaspectratio]{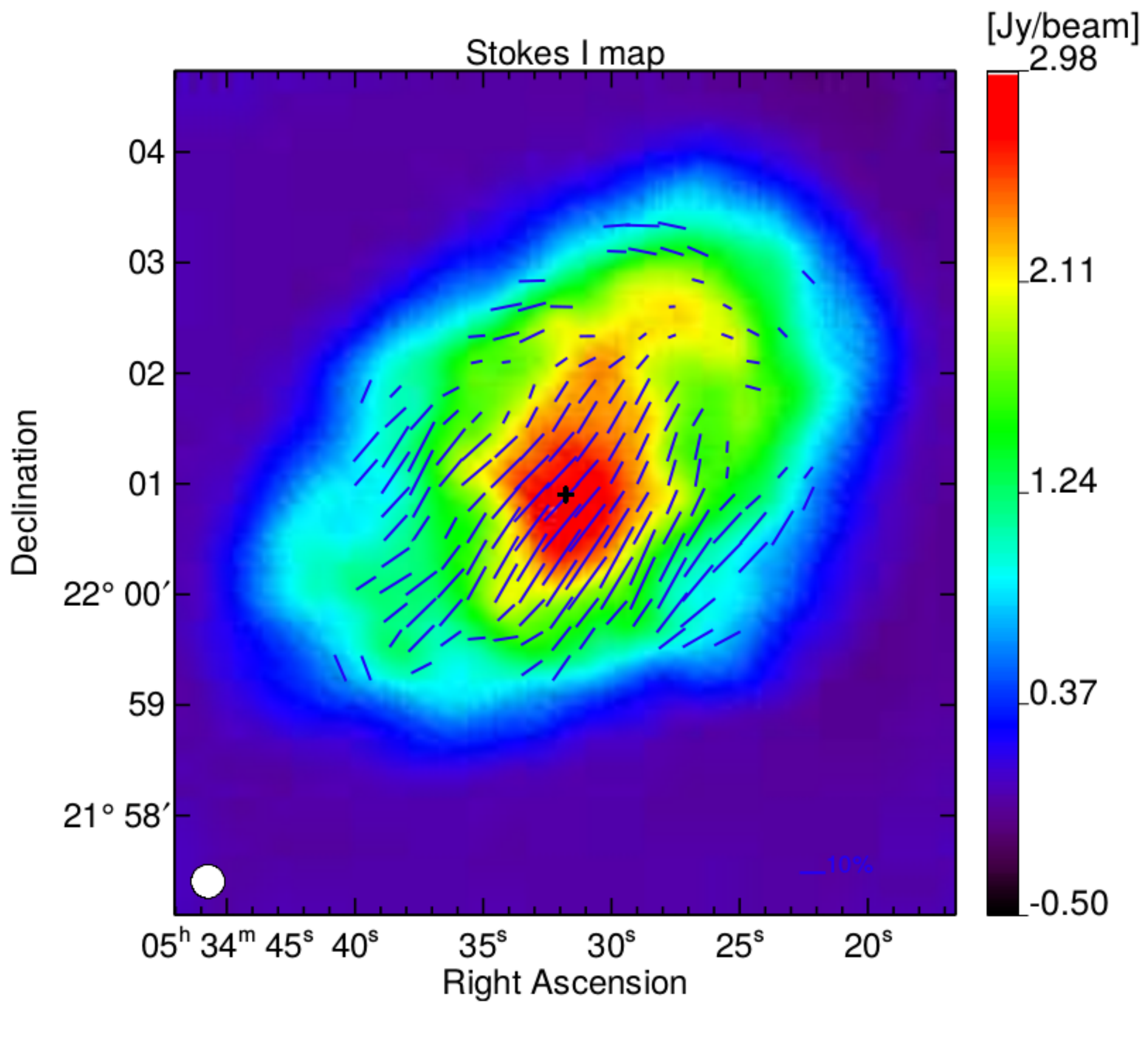}}      
             { \includegraphics[width=0.32\linewidth,keepaspectratio]{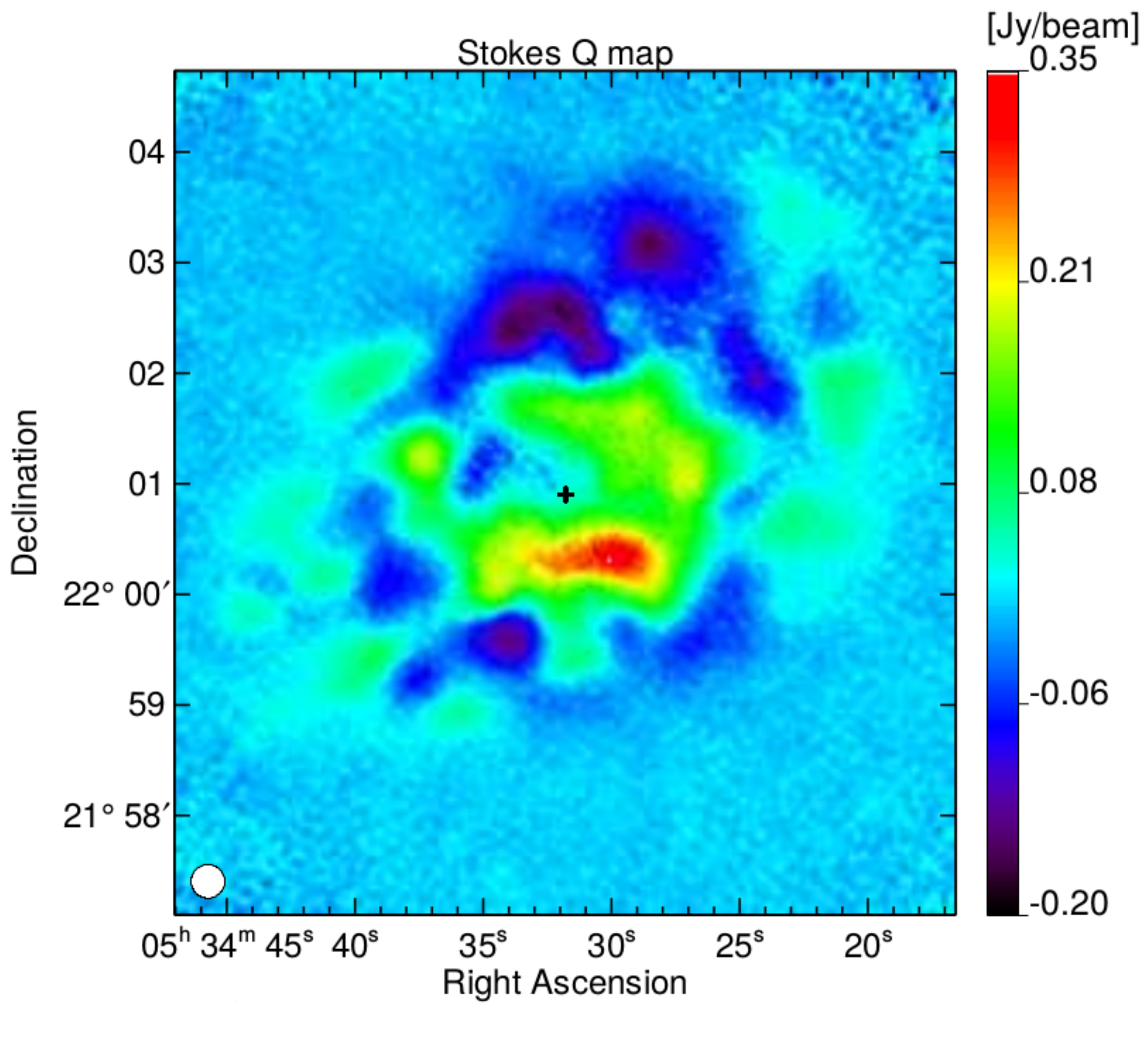}}
          { \includegraphics[width=0.32\linewidth,keepaspectratio]{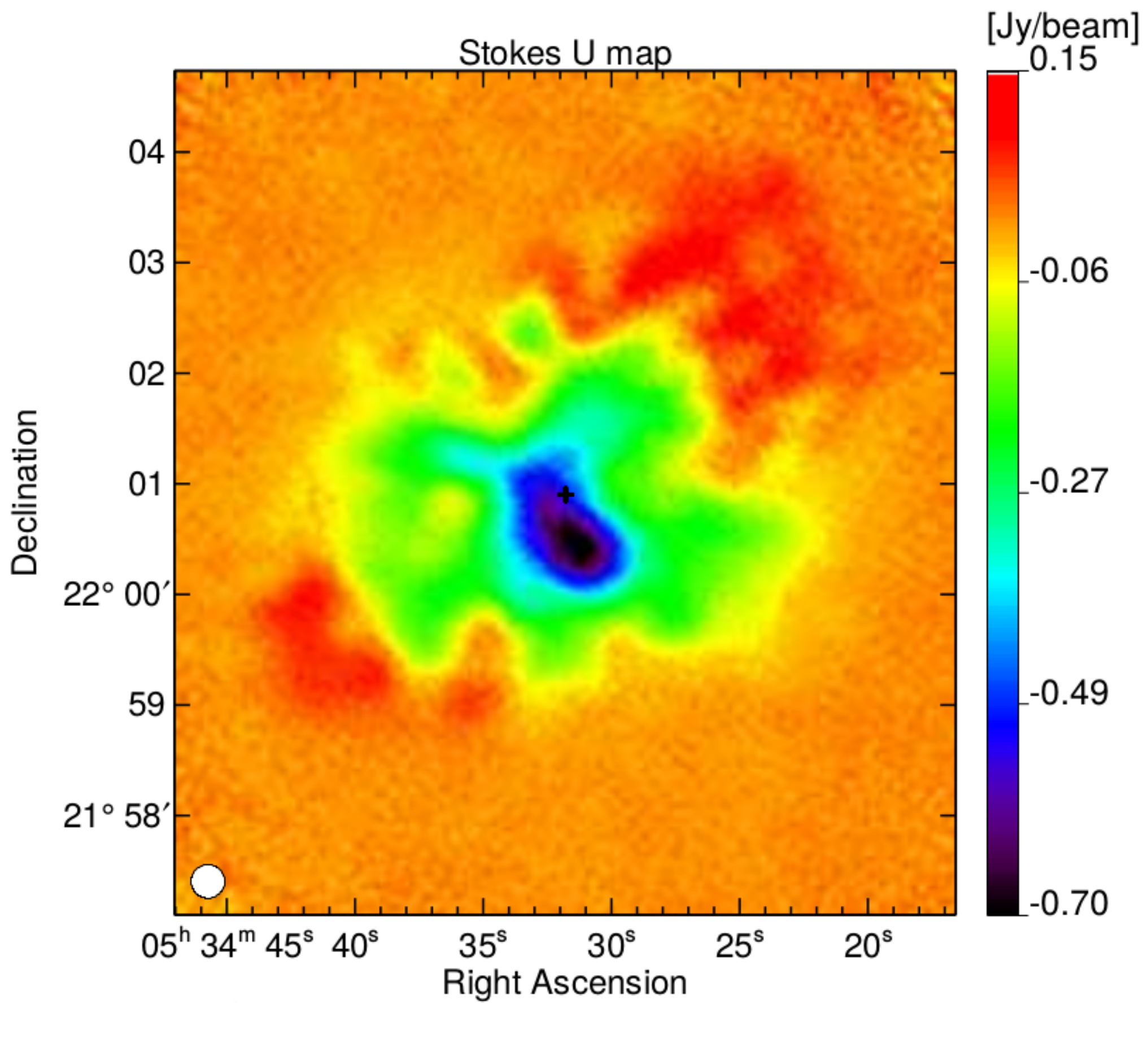}} 
           \caption{From left to right: \object{Crab} nebula Stokes $I$, $Q$, and $U$
             maps shown here in equatorial coordinates obtained at 150 GHz with the \NIKA\ camera. Polarization
             vectors, indicating both the polarization degree and the orientation, are
             overplotted in blue on the intensity map, where the polarization
             intensity satisfies $\textrm{I}_\textrm{pol} > 3 \sigma_{\textrm{I}_\textrm{pol}}$ and $\textrm{I}_\textrm{pol} >$ 0.1 Jy/beam. In each map the \NIKA\ FWHM is shown as a white disk in the bottom left corner and the cross marks the pulsar position.}
\label{crab_intensity_maps}
\end{figure*}

\section{Introduction}\label{sec:introduction}
The \object{Crab} nebula (or Tau A) is a plerion-type supernova remnant emitting a highly
polarized signal \citep{1978A&A....70..419W,1991ApJ...368..463M}.
Referring to \citet{2008ARA&A..46..127H}, from inside out the \object{Crab} consists of a pulsar, the synchrotron nebula, a bright expanding shell of thermal gas, and a larger very faint freely expanding supernova remnant.
Near the centre  of the nebula a shock is observed; it is formed by the jet's thermalized wind, which is confined by the thermal ejecta from the explosion \citep{2000ApJ...536L..81W,2011A&A...528A..11W}.
The synchrotron emission from the nebula is observed in the radio frequency domain and is  powered by the pulsar located at equatorial coordinates (J2000) $R.A. = 5^h34^m31.9383014s$ and $Dec. = 22^{\circ}0^{\prime}52.17577^{\prime\prime}$ \citep{Lobanov} through its jet.
The polarization of the \object{Crab} nebula radio emission, discovered in 1957 independently by \citet{1957ApJ...126..468M} and \citet{1959SvA.....3...39K}, has confirmed that the synchrotron emission is the underlying mechanism. 

Today the \object{Crab} nebula is perhaps the most observed object in the sky beyond our own solar system \citep{2008ARA&A..46..127H} and often used as a calibrator by new instruments. It is also quite isolated
with low background diffuse emission. In particular, it is the most intense polarized astrophysical object in the microwave sky
at angular scales of a few arcminutes and for this reason it is chosen not only for high resolution cameras,
but also for the calibration of
cosmic microwave background (CMB) polarization experiments, which have
beamwidths comparable to the extension of the source. 
Upcoming CMB experiments aiming at measuring the primordial $B-$modes require an accurate
determination of the foreground emissions to the CMB signal and a high control
of systematic effects.
The \object{Crab} nebula has already been used for polarization cross-check analysis in the frequency
range from 30 to 353 GHz \citep{2011ApJS..192...19W,2015arXiv150702058P}.
High angular resolution observations from the XPOL experiment \citep{thum2008} at the
IRAM 30 m telescope have revealed the spatial distribution of the \object{Crab} nebula  in
total intensity and polarization at 90 GHz with an absolute accuracy of 0.5$^{\circ}$
in the polarization angle \citep{aumont2010}.
This observation has also shown that the polarization spatial distribution varies from the source peak
to the edges of the source, and illustrates the need of an accurate study at high resolution in a large frequency range to be able to use this source as a calibrator for future polarization esperiments.

Previous studies \citep{macias2010} of the total spectral energy distribution
(SED) have shown a spectrum well described by a single
synchrotron component at radio and millimetre wavelengths, and predict negligible
variations in polarization fraction and angle in the frequency range of interest
for CMB studies.
 
Observations of the \object{Crab} nebula polarization were performed with the \NIKA\ camera   \citep{monfardini2010,catalano2014,monfardini2014} at the IRAM 30
m telescope  during the observational campaign of February 2015. A first
overview of the  \NIKA/30m (hereafter NIKA)   \object{Crab} polarization observations,
focusing on instrumental characterization of the polarization system, was given in
\cite{2016JLTP..184..724R}. In this paper we go a step further in the analysis
by combining \NIKA\ observations with published values at other frequencies, and spanning different angular resolutions, to trace the
SED in total intensity and polarization of the \object{Crab} nebula. To date, for polarization we have used observations from the \WMAP\
satellite at 23, 33, 41, 61, and 94 GHz \citep{2011ApJS..192...19W}, XPOL/30m (hereafter XPOL) at 90 GHz
\citep{aumont2010}, POLKA/APEX (hereafter POLKA) at 345 GHz \citep{2014PASP..126.1027W}, and from the
\Planck\ satellite at 30, 44, 70 \citep{2015arXiv150702058P}, 100, 143, 217, 353 GHz to be published \citep{planck2018}.

The paper is organized as follows: in Sect.~\ref{sec:NIKA observations} the
intensity and polarization maps obtained with the \NIKA\ camera are presented
together with the polarization degree and angle spatial distributions;
Sect.~\ref{sec:Polarization estimates in CMB experiments like beams} presents the
reconstruction of the polarization properties of the \object{Crab} nebula in well-defined
regions; Sect.~\ref{sec:Polarization intensity Spectral Energy Density (SED)}
presents the \object{Crab} nebula SED in total intenstity and polarization; and in
Sect.~\ref{sec:conclusions} we present our conclusions.
 
\section{\NIKA\ observations of the Crab nebula}\label{sec:NIKA observations}
\subsection{\NIKA\ camera and polarization set-up}\label{sec:nika camera}
\NIKA\ is a dual-band camera observing the sky in total intensity and polarization at
150 and 260 GHz with 18~arcsec and 12~arcsec FWHM resolution, respectively. It
has a field of view (FoV) of 1.8$^{\prime}$ at both wavelengths. It was operated at the
IRAM 30~m telescope between 2012 and 2015. A detailed description of the
\NIKA\ camera can be found in \citet{monfardini2010, monfardini2011} and
\citet{catalano2014}.

In addition to total power observations, \NIKA\ was also a test
bench for the polarization system of the final instrument
\NIKAd\ \citep{calvo2016,catalano2016nika2,2017arXiv170700908A}, which was installed at the
telescope in October 2015. The polarization set-up of \NIKA\
consists of a continuously rotating metal mesh half-wave plate (HWP)
followed by an analyser, both at room temperature and placed in
front of the entrance window of the cryostat. The \NIKA\ Lumped Elements Kinetic
Inductance Detectors (LEKIDs) are not intrinsically sensitive to
polarization. The HWP rotates at 2.98 Hz allowing a modulation of the polarization signal at $4\times 2.98$~Hz, while the typical telescope scanning speed is equal to 26.23 arcsec/s.
These conditions provide a quasi-simultaneous measure of Stokes parameters $I$, $Q$, and $U$
per beam and place the polarized power in the frequency domain far from the low frequency
electronic noise and the atmospheric fluctuations. \cite{ritacco2017} gives more
details on the \NIKA\ polarization capabilities and describes the performance of
the instrument at the telescope. In particular the sensitivity of the
\NIKA\ camera in polarization mode was estimated to be 50 mJy.$s^{1/2}$ at 150
GHz.

\NIKA\ has provided the first polarization
observations performed with kinetic inductance detectors (KIDs), confirming that KIDs are also a
suitable detector technology for the development  of the next generation of polarization sensitive
experiments.

\subsection{\NIKA\ observations}\label{sec:nika_observations}
Observations of the polarized emission from the \object{Crab} nebula with the \NIKA\ camera were performed at
the IRAM 30~m telescope in February 2015. The average opacity at 150 GHz was $\tau$ = 0.2.  Figure~\ref{crab_intensity_maps} shows
the Stokes $I$, $Q$, and $U$ maps obtained by a co-addition of 14 maps
of $8 \times 6$ arcminutes for a total observation time of $\sim$ 2.4 hours. The rms calculated on jack-knife noise maps is 36 mJy/beam on Stokes $I$ maps and 31 mJy/beam on  Stokes $Q$ and $U$ maps.
The maps were performed in equatorial coordinates in four different scan
directions: 0$^{\circ}$, 90$^{\circ}$, 120$^{\circ}$, 150$^{\circ}$. This allowed us to have the best mapping with different position angles.

\begin{figure}[h!]
  \centering
  \includegraphics[width=1\linewidth,keepaspectratio]{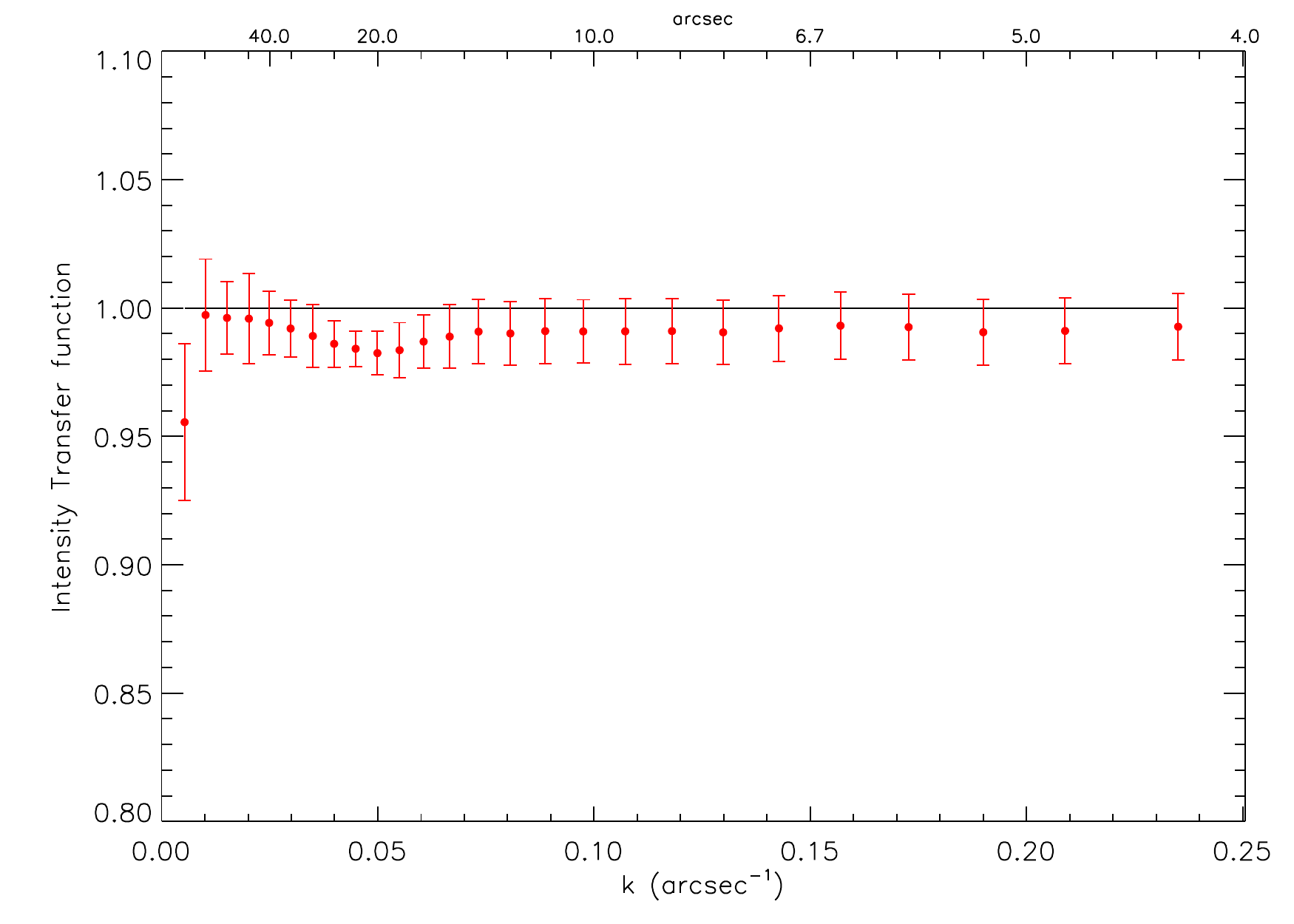}
        \caption{Transfer function of the data processing in total intensity.}
\label{transfer_func}
\end{figure}

To obtain the Stokes $I$, $Q$, and $U$ \object{Crab} nebula maps in  equatorial coordinates, we  used a dedicated
polarization data reduction pipeline \citep{ritacco2017}, which is an extension
of the total intensity \NIKA\ pipeline \citep{catalano2014,adam2013}. The main steps
of the polarization pipeline are summarized below:
\begin{enumerate}
\item Subtraction of the HWP induced parasitic signal, which is modulated at harmonics of the HWP rotation frequency and represents the most detrimental noise contributing to the polarized signal; 
\item Reconstruction of the Stokes $I$, $Q$, and $U$ time ordered information
  (TOI) from the raw modulated data. This is achieved using a demodulation
  procedure consisting in a lock-in around the fourth harmonic of the HWP rotation frequency where the polarization signal is located;
\item Subtraction of the atmospheric emission in the demodulated TOIs using
  decorrelation algorithms. In polarization, the HWP modulation 
  significantly reduces the atmospheric contamination and there is no need to
  further decorrelate the $Q$ and $U$ TOIs from residual atmosphere. By contrast, in
  intensity the atmospheric emission fully dominates the signal; to recover
  the large angular scales, we used the 260~GHz band as an atmosphere
  dominated band, as in \cite{adam2013}.  This decorrelation impacts the  reconstructed Stokes maps via a transfer function. We have estimated this function
  with simulated observations of diffuse emission that were passed through the
  data reduction pipeline, with the exact same scanning, sample flagging, and data
  processing as real data. We found that the power spectrum of the output map
  matches that of the input map to better than 1\% (~5\%) on scales smaller  (larger) than $\sim 1'$, see Fig.~\ref{transfer_func}. Its moderate effect on large angular scales is further reduced with the subtraction of a zero level for the photometry (see below), so the impact of the data processing is thus
  negligible compared to uncertainties on absolute calibration on small
  scales. In the
  following, we therefore neglect the impact of this transfer function.

\item Correction of the intensity-to-polarization leakage effect, which was
  identified in observations of unpolarized sources like the planet Uranus. For
  point sources the effect was about 3\% peak to peak, while for extended sources like the \object{Crab} nebula it has been found to be of the order of 0.5\% peak to peak. 
  \cite{ritacco2017} describe the algorithm of leakage correction developed specifically for \NIKA\ polarization observations. Applying this algorithm to Uranus observations the instrumental polarization is reduced to 0.6\% of the total intensity $I$;
  \item Projection of the demodulated and decorrelated Stokes $I$, $Q$, and $U$ TOIs into Stokes $I$, $Q$, and $U$ equatorial coordinates maps.

\end{enumerate}

  \begin{figure*}
\centering
\includegraphics[clip, angle=0, scale = 0.35]{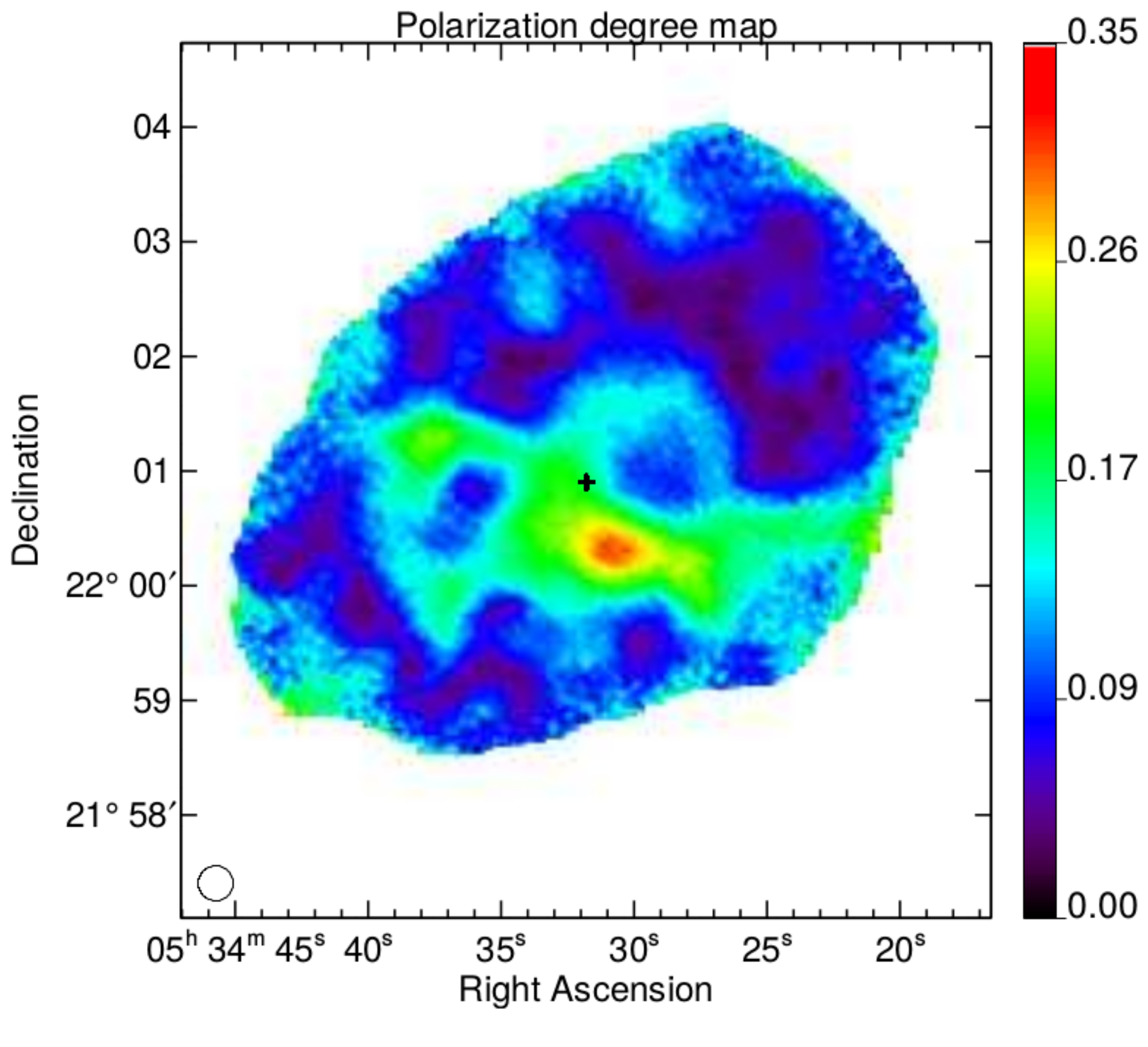}
\includegraphics[clip, angle=0, scale = 0.5]{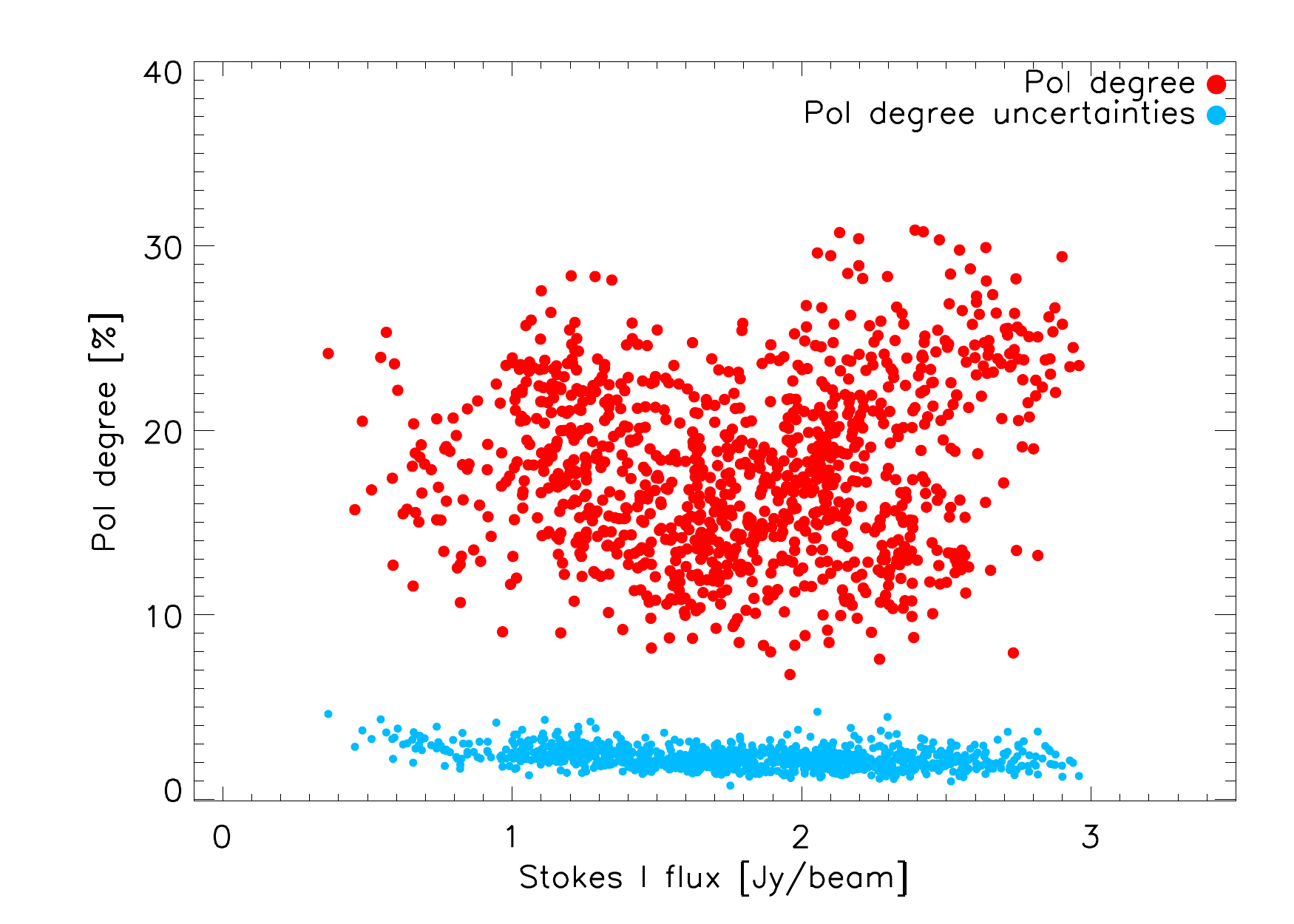}
\includegraphics[clip, angle=0, scale = 0.35]{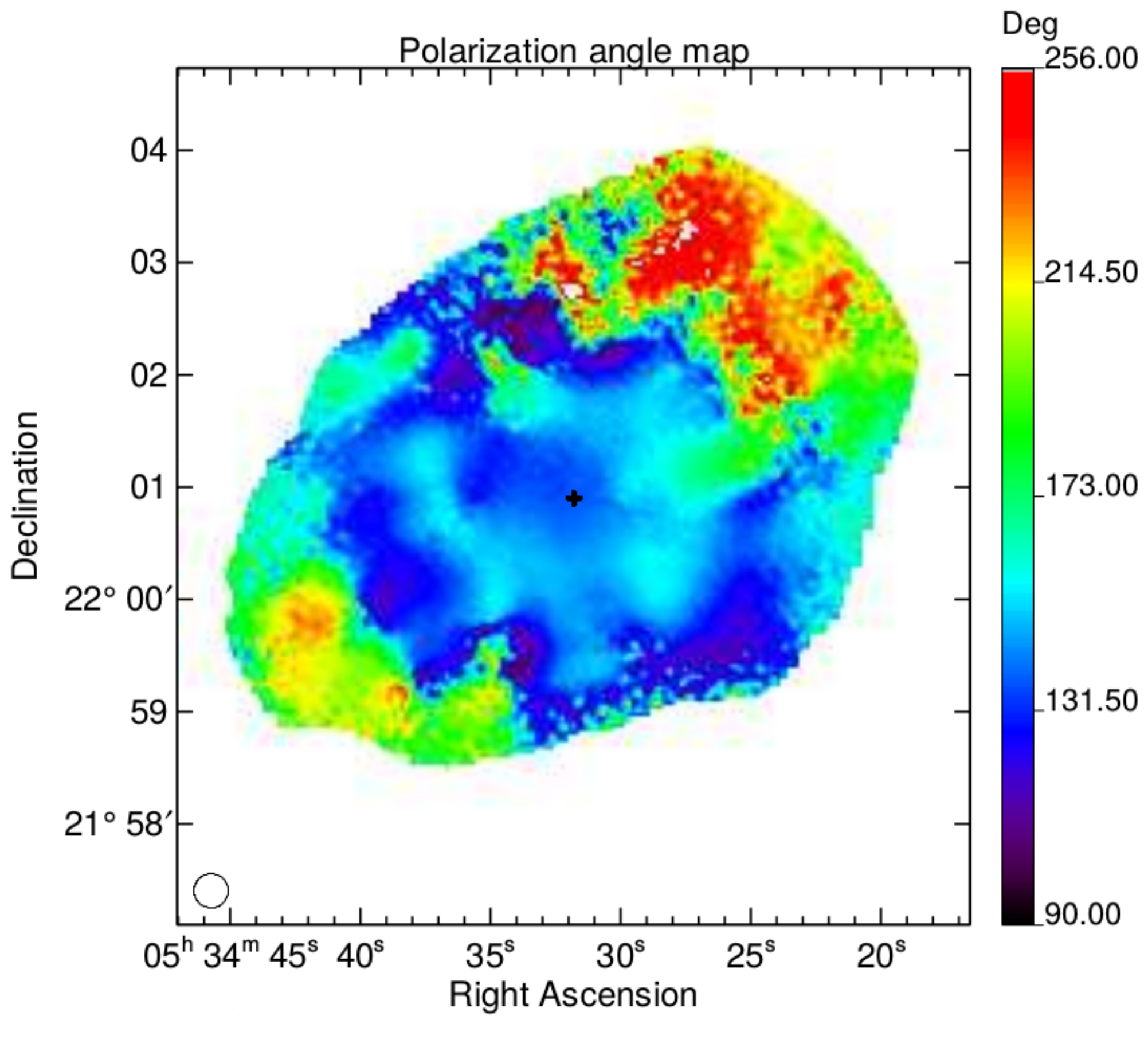}
\includegraphics[clip, angle=0, scale = 0.5]{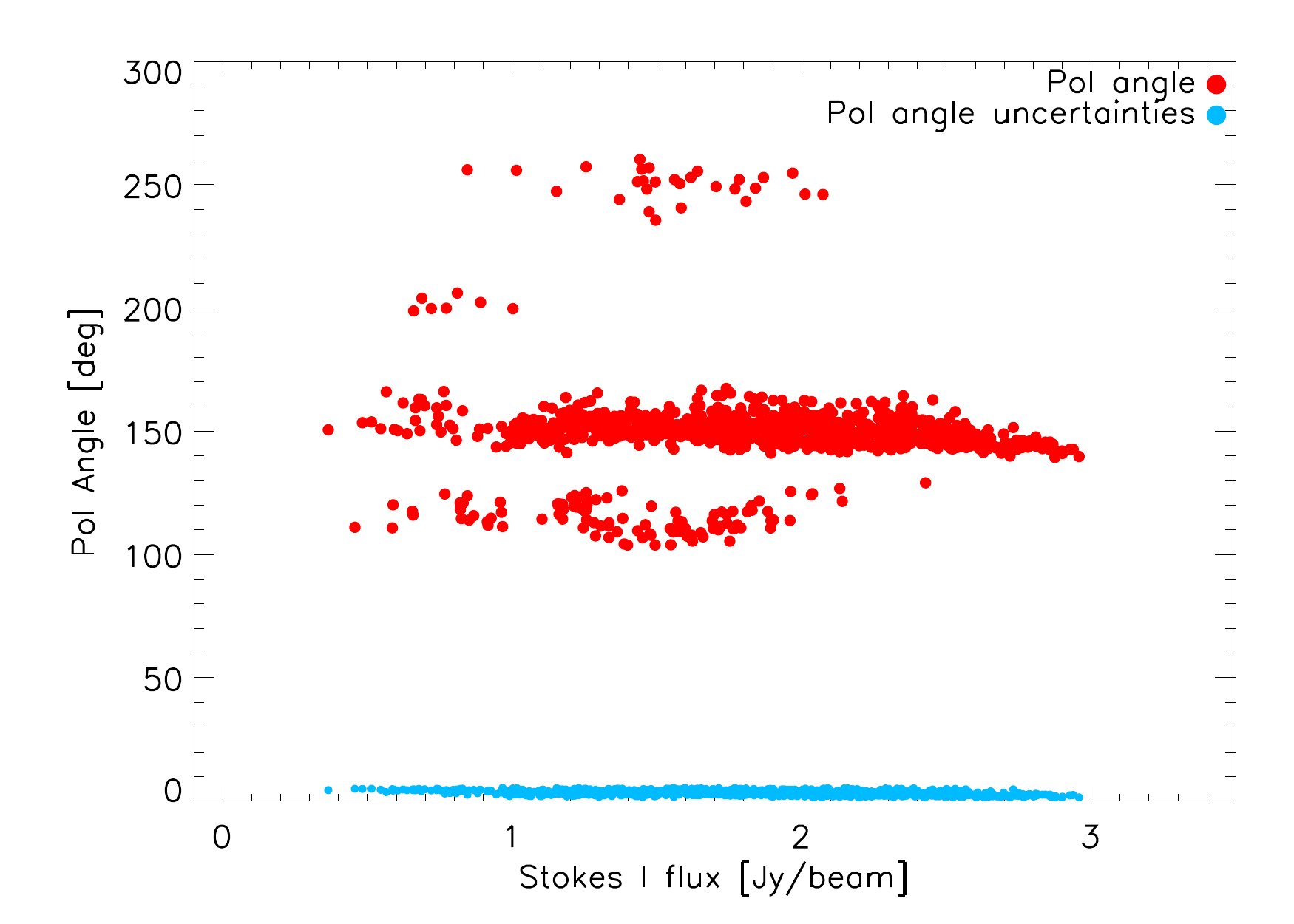}
\caption{{\it Top left}: 
Polarization degree map $p$, uncorrected for noise bias, of the \object{Crab} nebula. {\it Top right}: Noise bias
  corrected $p$ values as a function of total intensity map (Stokes $I$). 
  The condition $\textrm{I}_\textrm{pol}$ $\textgreater$ 5$\sigma_{\textrm{I}_\textrm{pol}}$ is satisfied for those values. {\it
    Bottom left}: Polarization angle map $\psi$ (equatorial coordinates system) of the
  \object{Crab} nebula. {\it Bottom right}: Distribution of $\psi$ values  represented as a function of the total intensity in the case of very high S/N  where
 $\textrm{I}_\textrm{pol}$ $\textgreater$ 5$\sigma_{\textrm{I}_\textrm{pol}}$. The cyan dots represent the uncertainties calculated as the dispersion between different observational scans. The black cross marks the pulsar position on the maps.
  }
\label{fig:pol_degree}
\end{figure*}
 \begin{figure}
  \centering
      {\includegraphics[width=0.75\linewidth,keepaspectratio]{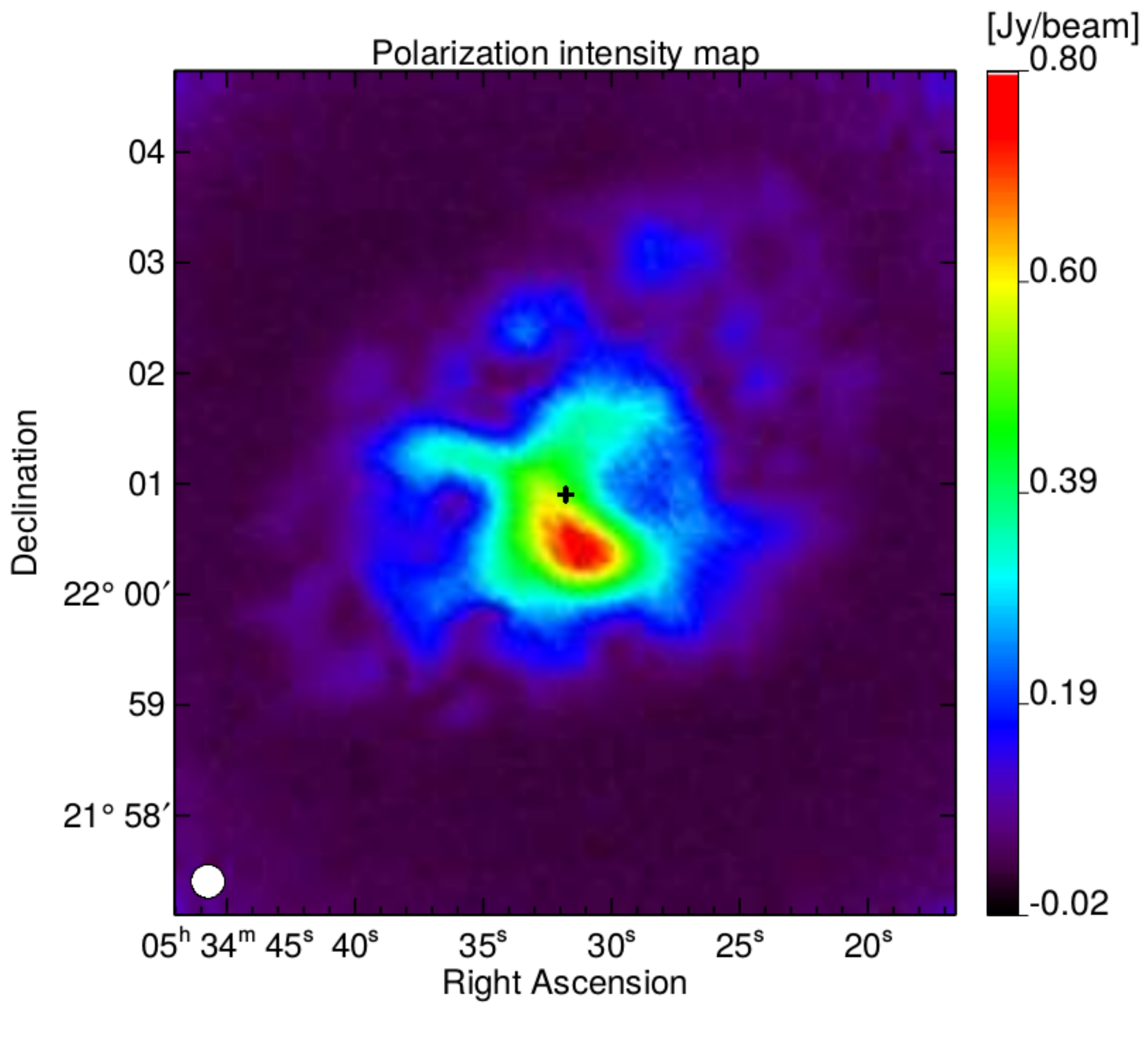}}
\caption{\NIKA\ polarized intensity map of the  \object{Crab} nebula at 150 GHz. The map shows high polarized emission reaching a value of 0.8 Jy beam$^{-1}$. The telescope beam FWHM is shown in the lower left. The black cross marks the pulsar position.}
\label{crab_ipol_maps}          
  \end{figure}

\subsection{Crab polarization properties}\label{sec:pol_properties}
In this section we discuss the polarization properties of the source in terms of polarization degree $p$ and angle $\psi$, which are defined through the Stokes parameters $I, Q$, and $U$ as follows:
\begin{equation}
 p    = \frac{\sqrt{Q^2 + U^2}}{I} \nonumber 
\end{equation}
and
 \begin{equation}
 \psi = \frac{1}{2}\arctan\frac{U}{Q}.\label{angledegree_polar}
 \end{equation}
The polarization angle follows the IAU convention, which counts east from north in the equatorial coordinate system.

These definitions are not linear in $I$, $Q$, and $U$, and therefore the observational uncertainties have to be carefully considered, {i.e.} $p$ and $\psi$ are noise biased. 
\citet{1980A&A....91...97S,1985A&A...142..100S} and \citet{montier} proposed analytical solutions to correct for this bias. For intermediate and high S/N ratio the polarization degree and its uncertainty read
 \begin{eqnarray}
 p    &=& \frac{\sqrt{Q^2 + U^2 - \sigma_{Q}^2 - \sigma_{U}^2}}{I}, \nonumber \\ 
  \sigma_{p} &=& \frac{\sqrt{Q^2\sigma_Q^2 + U^2\sigma_U^2 + p^4I^2\sigma_I^2}}{pI^2}.
  \label{p_true_degree}
 \end{eqnarray}
Furthermore, the polarization angle in a high S/N regime can be approximated by Eq.~\ref{angledegree_polar} with the uncertainty
 
  \begin{eqnarray}\label{angle_uncertainty}
  \sigma_{\psi} = \frac{\sqrt{Q^2\sigma_U^2 + U^2\sigma_Q^2}}{2(pI)^2}.  \end{eqnarray}

The spatial distribution map of the polarization degree $p$ of the \object{Crab} nebula
without noise bias correction is presented in the top left panel of
Fig.~\ref{fig:pol_degree}. We note that we have set to 0 the pixels for which the total intensity is lower than 0.1 Jy/beam to avoid the divergence of $p$. The same has been done for the polarization angle map (bottom left of Fig.~\ref{fig:pol_degree}.)

The polarization degree $p$ reaches a value of 20.3$\pm$0.7\% at
the peak of the total intensity, which is consistent with what is observed in the
top right panel of Fig.~\ref{fig:pol_degree} where the variation of $p$ as a
function of the Stokes $I$ is shown.  Here the $p$ values have been noise bias
corrected and satisfy the condition $\textrm{I}_{\textrm{pol}}=\sqrt{Q^2+U^2}$ $\textgreater$ 5
$\sigma_{\textrm{I}_{\textrm{pol}}}$. The distribution of the polarization degree appears highly
dispersed around a mean value of 20\%. These points are mostly located around the peak of Stokes $I$. 
The spatial variation of $p$ highlights the interest of high resolution polarization
observations of the \object{Crab} nebula. 

The bottom left panel of Fig.~\ref{fig:pol_degree} shows the spatial distribution of polarization angle
$\psi$. As discussed in \cite{ritacco2017} a 1.8$^{\circ}$ uncertainty must be considered in the polarization angle coming from the determination of the HWP zero position, corresponding to its optical axis in the \NIKA\ cabin reference frame. An uncertainty of 0.5$^{\circ}$ must be considered due to the leakage effect subtraction, which has been estimated from the comparison of the maps before and after leakage correction. We observe a relatively constant polarization
angle 140$^{\circ}$ $\textless$ $\psi_{\textrm{eq}}$ $\textless$150$^{\circ}$ represented here in equatorial coordinates, except
for the north-east region where the averaged angle is around $250^{\circ}$, and
some inner regions with lower polarization angle.  These values are confirmed by
the bottom right panel, which shows the polarization angle distribution as a
function of total intensity satisfying the condition ${\textrm{I}_{\textrm{pol}}} > 5\sigma_{{\textrm{I}_{\textrm{pol}}}}$.

The sudden change in polarization angle on the northern region was already
observed by the XPOL experiment at 90 GHz \citep{aumont2010}.  This together
with the variation in the polarization fraction discussed above confirms the
need of high angular resolution observations at low and high frequencies for a
good understanding of the \object{Crab} polarized emission properties.
High resolution observations give the possibility to estimate the polarization properties at different scales and to make a comparison  with low resolution experiments, for example CMB experiments.

We present in Fig.~\ref{crab_ipol_maps} the 150 GHz \object{Crab} polarization intensity
map ${\textrm{I}_{\textrm{pol}}}$ uncorrected for noise bias. We observe a peak at 0.8 Jy beam$^{-1}$ and the polarization
decreases towards the edges of the nebula. 

\subsection{Comparison to other high resolution experiments}
Following previous studies, we compare here the \NIKA\ results at the pulsar position and at Stokes $I$ map peak with high angular resolution experiments such as POLKA, XPOL, and SCUPOL/JCMT (hereafter SCUPOL). These experiments observed at wavelengths of 870 $\mu$m, 3 mm, and 850 $\mu$m, respectively.
In order to compute polarization estimates in the same region around these positions, we use POLKA \citep{2014PASP..126.1027W}, XPOL \citep{aumont2010}, SCUPOL \citep{scubapol}, and \NIKA\ (this paper) maps degraded at an XPOL angular resolution of 27$^{\prime\prime}$. The region is defined as an XPOL pixel size of 13.7$^{\prime\prime}$.

The results obtained are shown in Table~\ref{tab:peak_pulsar_others}. The polarization degree $p$ values are consistent for all the instruments at both positions, while for the angle $\psi_{eq}$ we observe a fair agreement. For POLKA and SCUPOL the values found are consistent with those in~\citet{2014PASP..126.1027W}.
For XPOL the values found at the pulsar position differ from those in~\citet{aumont2010}. We note that it is hard to estimate  the polarization at the pulsar position precisely because it is located where the polarization changes drastically, while the peak of the total intensity is very well defined. 

\begin{table}[h!]
  \centering
      \begin{tabular}{ccccccccc}
     &  & \small $p$ [\%] & \small $\psi_{\textrm{eq}}$ [$^\circ$] & \\ 
\hline
\hline
      &\small POLKA  & \small 18.2 $\pm$ \small 4.8 & \small 147.1 $\pm$ \small 7.5  \\
      \small Pulsar &\small XPOL   & \small 17.5 $\pm$ \small 1.2 & \small 150.2 $\pm$ \small 2.0   \\
     &\small SCUPOL & \small 14.8 $\pm$ \small 2.8 & \small 143.5 $\pm$ \small 4.4 & \\
      &\small \NIKA\ & \small 17.9 $\pm$ \small 2.2 & \small 138.8 $\pm$ \small 1.5 $\pm$ \small 2.3 \\
      \hline

      &\small POLKA  & \small 19.4 $\pm$ \small 4.4 & \small 148.1 $\pm$ \small 6.5   \\
       \small Intensity Peak &\small XPOL  &  \small 21.0 $\pm$ 1.2 & \small 149.0 $\pm $ \small 1.6   \\
      &\small SCUPOL  & \small 16.4 $\pm$ 4.8 & \small 151.8 $\pm$ \small 8.4\\
      &\small \NIKA\ & \small 20.3 $\pm$ \small 0.7 & \small 140.0 $\pm$ \small 1.0 $\pm$ \small 2.3 \\
\hline
\hline
    \end{tabular}
   \caption{Polarization degree and angle estimated around the pulsar and intensity peak positions for POLKA \citep{2014PASP..126.1027W}, XPOL \citep{aumont2010}, SCUPOL \citep{scubapol}, and \NIKA\ (this paper). The values have been estimated using the maps and degrading  them to the XPOL angular resolution of 27$^{\prime\prime}$. For \NIKA\ the position of the pulsar, represented on the maps by a black cross, refers to \cite{Lobanov}. The position of the peak of the total intensity measured on the \NIKA\ maps has equatorial coordinates (J2000) $R.A. =5^h34^m32.3804s$ and $Dec. = 22^{\circ}0^{\prime}44.0982^{\prime\prime}$. The polarization angle is given here in equatorial coordinates. A systematic angle uncertainty of 2.3$^{\circ}$ is considered. A total calibration error of 10 $\%$ has been accounted for in the flux estimates.}
\label{tab:peak_pulsar_others}
 \end{table}

\section{Total intensity and polarization fluxes}\label{sec:Polarization estimates in CMB experiments like beams}
\subsection{Total intensity flux}
\begin{figure}[h!]
  \centering
  \includegraphics[width=0.7\linewidth,keepaspectratio]{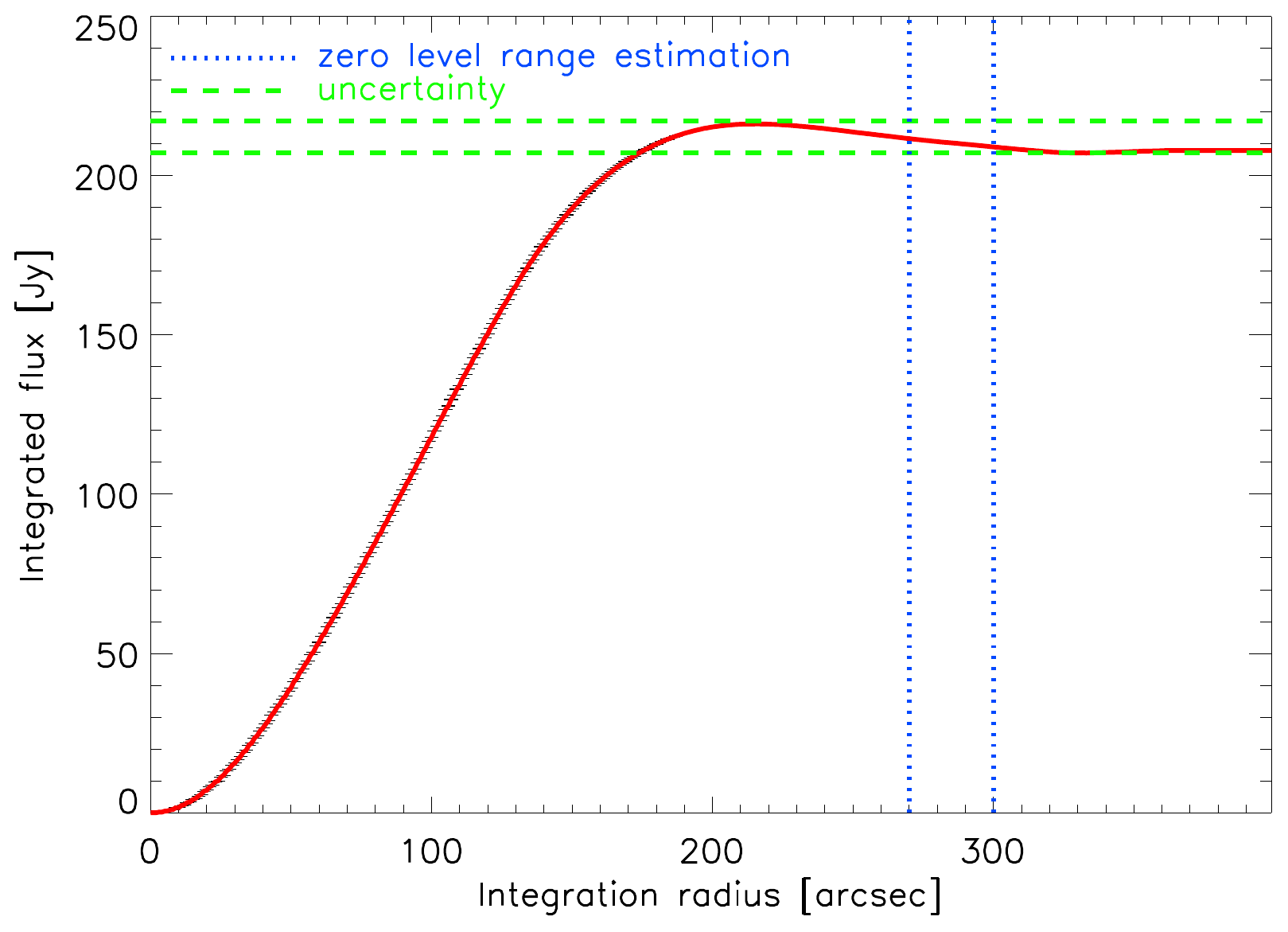}
  \includegraphics[width=0.8\linewidth,keepaspectratio]{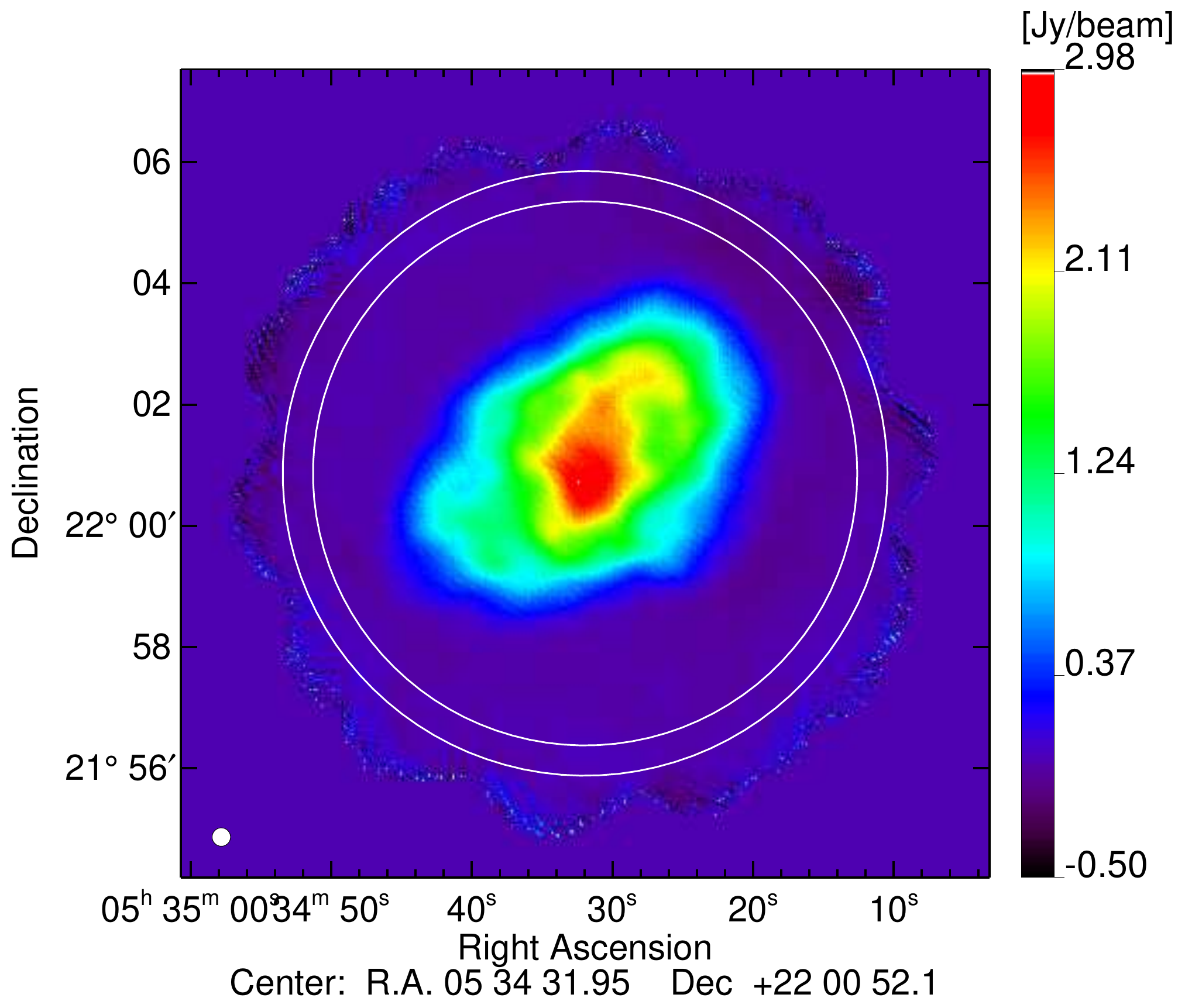}
     \caption{
       Cumulative flux of the \object{Crab} nebula (top) obtained at 150 GHz over
       5$^{\prime}$ from the centre obtained by aperture photometry. The flux
       has been corrected by a zero level in the map, which corresponds to the
       mean of the signal calculated in an annular ring, as indicated by the
       white circles on the map and by the blue dotted lines on the
       top. The green dotted line represents the uncertainties measured at large
       radii.}
\label{crab_integrated_flux}
\end{figure}

We computed the total flux across the \object{Crab} nebula, which has an extent of about
5$^{\prime}$x7$^{\prime}$ (see Fig.~\ref{crab_intensity_maps}).  We used
standard aperture photometry techniques to calculate the flux as shown in the top panel of
Fig.~\ref{crab_integrated_flux}. We used as centre position the centre of the
map with equatorial coordinates (J2000) $R.A. = 5^h34^m31.95s$ and $Dec. = 22^{\circ}0^{\prime}52.1^{\prime\prime}$. A zero level in the map, calculated as the mean of the signal measured on
an external annular ring region (see bottom panel of
Fig.~\ref{crab_integrated_flux}) of radius 4.5$^\prime$ $\textless$ R $\textless$
5$^\prime$, has been subtracted from the map. The total signal estimated is
222.7$\pm$1.0$\pm$1.3$\pm$22.4 Jy. The first uncertainty term accounts for statistical
uncertainties computed from fluctuations of the signal at large radii. The second uncertainty accounts for the difference between two sets of jack-knife noise maps. The third one accounts for the absolute calibration error of 10\%. A colour correction factor of 1.05 has been accounted for. This factor was estimated using the spectral index $\beta$ measured by \citet{macias2010} and considering the \NIKA\ bandpass at 2 mm, as shown in \citet{catalano2014}. The final flux is thus the integrated value in the bandpass.
We use Uranus for absolute point source flux calibration. The flux of the planet is estimated from a frequency dependent model of the planet brightness temperature as described in \cite{moreno2010}. 
This model is integrated over the \NIKA\ bandpasses for each channel, and it is assumed to be accurate at the 5\% level. The final absolute calibration factor is obtained by fitting the amplitude of a Gaussian function of fixed angular size on the reconstructed maps of Uranus, which represents the main beam. For the polarization observational campaign of February 2015 this uncertainty is estimated to be 5\% for the \NIKA\ 2.05 mm channel (150 GHz) \citep{ritacco2017}. 
Nevertheless, as described in \cite{adam2013} and \cite{catalano2014}, by integrating the Uranus flux up to 100 arcsec, we observe that the total solid angle covered by the beam is larger than the Gaussian best fit of the main beam by a factor of 28\%. As a consequence we account for this factor in the estimation of the fluxes.
Moreover, \cite{adam2013} estimated the uncertainty on the solid angle of the main beam to be 4\%.
Finally, the overall calibration error is estimated to be about 10\% by also considering  the uncertainties on the side lobes.

\begin{figure}
  \centering
          { \includegraphics[width=1\linewidth,keepaspectratio]{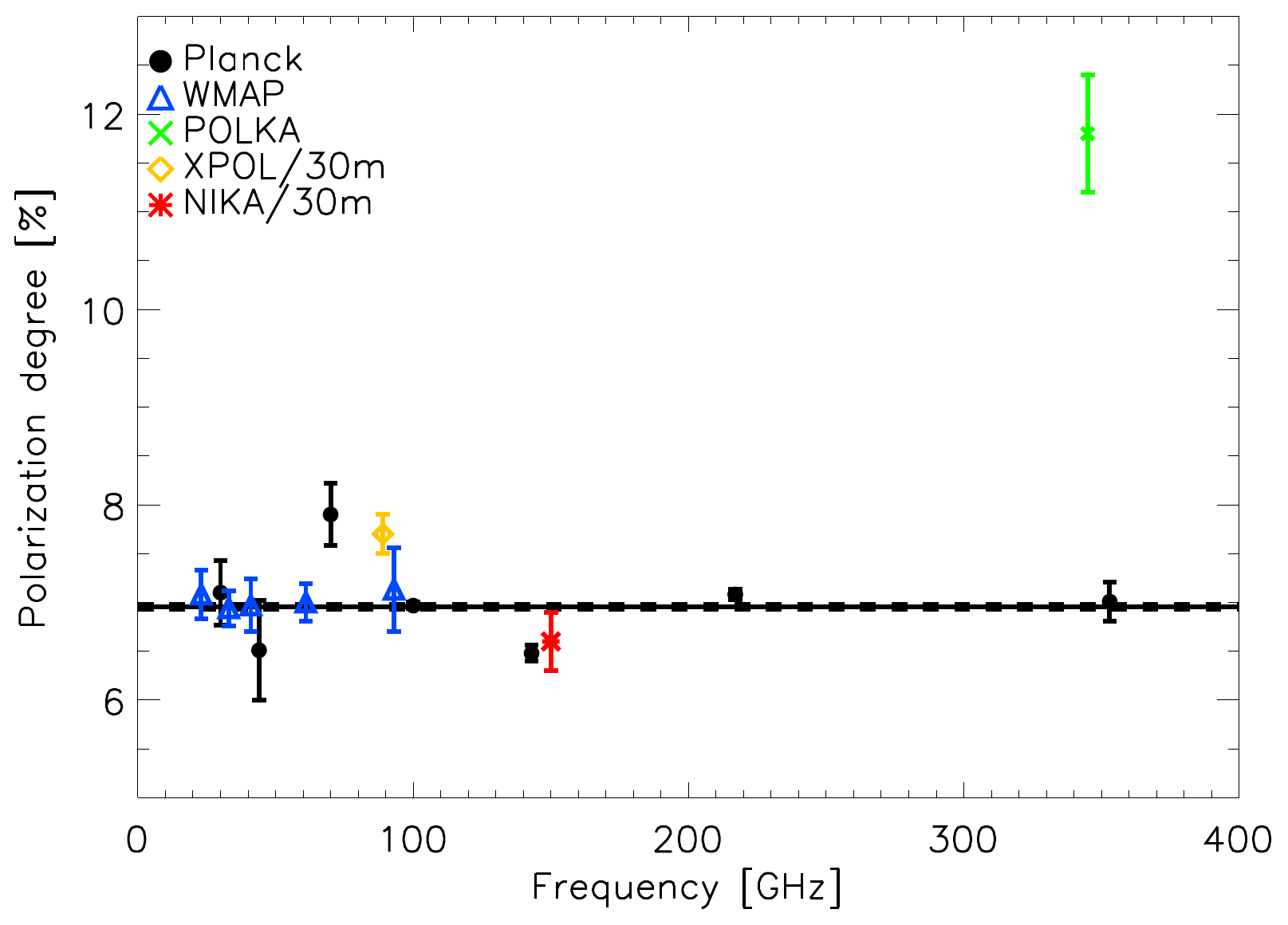}}
          { \includegraphics[width=1\linewidth,keepaspectratio]{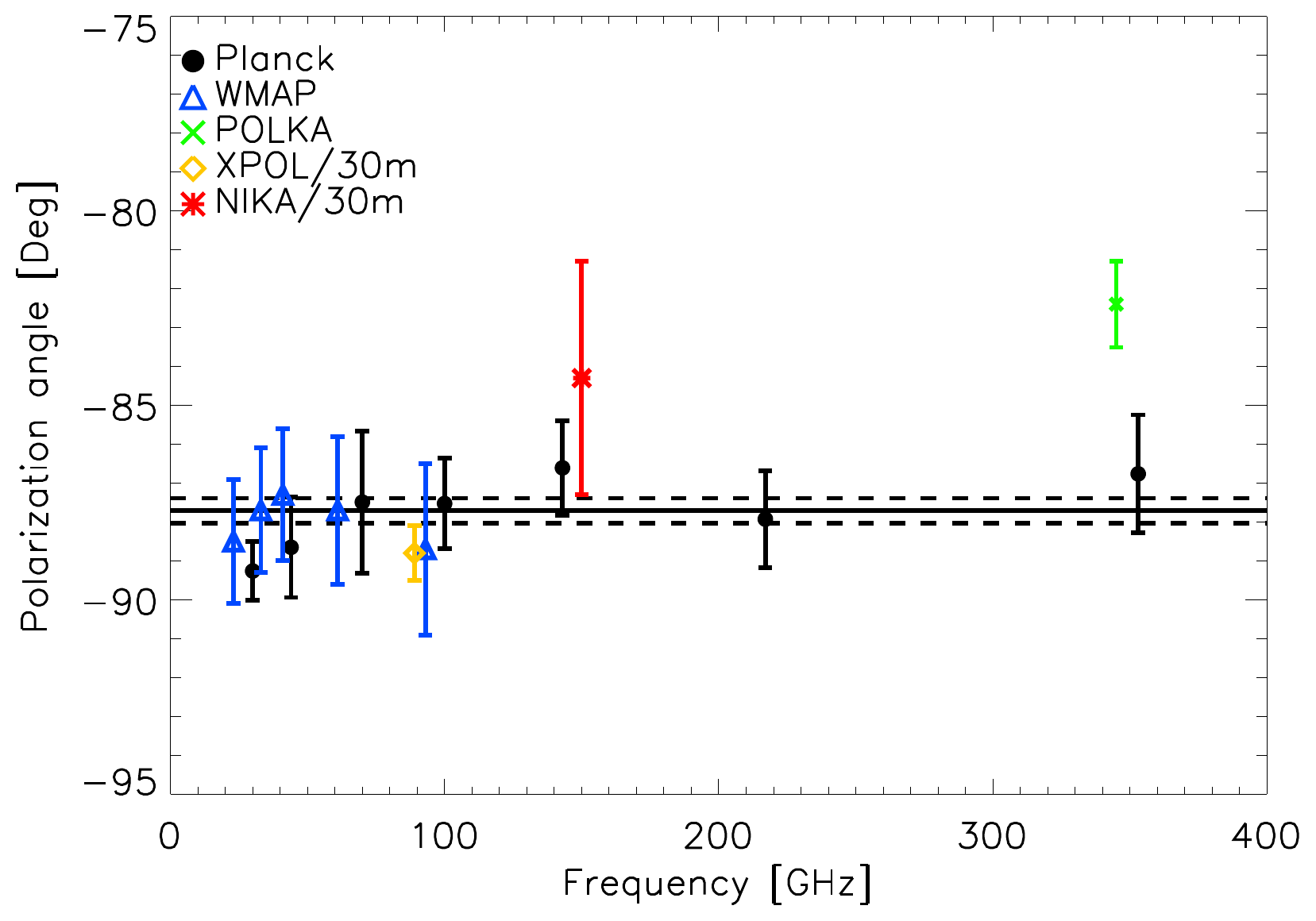}} 
            \caption{{\it Top}: Polarization degree as a function of frequency as measured by \Planck\ (black dots), WMAP (blue triangles), XPOL (yellow diamond), POLKA (green cross), and \NIKA\ (red crosses). The \NIKA\ and POLKA values have been estimated by aperture photometry considering the source extension up to $9^{\prime}$.  \Planck\ and WMAP values are shown at their native resolution. XPOL, \NIKA,\ and POLKA values have been integrated over the source. The solid line represents the weighted-averaged degree for all experiments but POLKA. Dashed lines represent 1$\sigma$ uncertainties.
            {\it Bottom}: Polarization angles in Galactic coordinates for the same five experiments. The solid line represents the weighted-averaged polarization angle computed using all the values.}
\label{crab_p_angle_comparison}         
  \end{figure}
\begin{table*}[h!]
  \centering
      \begin{tabular}{ccccccccc}
      \hline
      \hline
       & \small $I$ [Jy] & \small $Q$ [Jy] & \small $U$ [Jy] & \small I$_\textrm{pol}$ [Jy] & \small $p$ [\%] & \small $\psi_{eq}$ ($\psi_{gal}$) [$^\circ$] \\
      \hline

\small 5$^{\prime}$ & \small 191.4$\pm$26.8 & \small 2.97$\pm$0.3 & \small -16.0$\pm$1.5 & \small 16.3$\pm$1.7 & \small 8.5$\pm$0.4 & \small 140.3 (-82.6)$\pm$0.1$\pm$0.5$\pm$1.8$^*$  \\ 
\small 7$^{\prime}$ & \small 226.5$\pm$25.0 & \small 3.5$\pm$0.4 & \small -14.9$\pm$1.2 & \small 15.3$\pm$1.8 & \small 6.7$\pm$0.1 & \small 141.7 (-84.1)$\pm$0.2$\pm$0.5$\pm$1.8$^*$ \\ 
\small 9$^{\prime}$ & \small 222.7$\pm$24.6 & \small 3.5$\pm$0.4 & \small -14.3$\pm$1.2 & \small 14.8$\pm$1.6 & \small 6.6$\pm$0.3 & \small 142.0 (-84.3)$\pm$0.7$\pm$0.5$\pm$1.8$^*$ \\ 
\hline            
    \hline   
    \end{tabular}
   \caption{ 
   \NIKA\ \object{Crab} nebula total flux $I$, $Q$, and $U$ are  presented here, estimated by using aperture photometry methods. A colour correction factor of 1.05 has been taken into account. Polarized intensity flux I$_\textrm{pol}$, polarization degree $p$, and angles $\psi_{eq}$ (equatorial coordinates) and $\psi_{gal}$ (Galactic coordinates in brackets), are also presented. The values have been calculated within 5$^{\prime}$, 7$^{\prime}$, 9$^{\prime}$ by aperture photometry.
   A total calibration error of 10 $\%$ has been accounted for. The statistical uncertainty also accounts  for Monte Carlo simulations of the noise in $Q$ and $U$ and the differences between two sets  of seven jack-knife maps.
 *A systematic angle uncertainty of 1.8$^{\circ}$ must be considered in the polarization angle error budget. We also consider a 0.5$^{\circ}$ of uncertainty due to the intensity to polarization leakage correction.    
    }
    \label{tab:crab_results}
 \end{table*}
 \begin{table*}[h!]
  \centering
      \begin{tabular}{ccccccccc}
      \hline
      \hline
       Frequency [GHz] & \small $I$ [Jy] & \small $Q$ [Jy] & \small $U$ [Jy] & \small I$_\textrm{pol}$ [Jy] & \small $p$ [\%] & \small $\psi_{gal}$ [$^\circ$] \\
      \hline

\small 100 & \small 229.23$\pm$1.15  & \small -15.92$\pm$0.07 & \small 1.38$\pm$0.09 & \small 15.99$\pm$0.15 & \small 6.97$\pm$0.03 & \small -87.52$\pm$0.16  \\ 
\small 143 & \small 193.21$\pm$2.67  & \small -12.43$\pm$0.14 & \small 1.48$\pm$0.09 & \small 12.52$\pm$0.29 & \small 6.48$\pm$0.08 & \small -86.61$\pm$0.21  \\
\small 217 & \small 172.73$\pm$1.60  & \small -12.20$\pm$0.08 & \small 0.88$\pm$0.11 & \small 12.23$\pm$0.17 & \small 
7.08$\pm$0.05 & \small -87.93$\pm$0.25  \\
\small 353 & \small 144.84$\pm$1.75  & \small -10.01$\pm$0.29 & \small 1.15$\pm$0.18 & \small 10.16$\pm$0.60 & \small 7.01$\pm$0.20 & \small -86.76$\pm$0.52 \\
    \hline            
    \hline   
    \end{tabular}
      \caption{The same quantities as in Table~\ref{tab:crab_results} derived with the new analysis of HFI \Planck\ results by using aperture photometry of the polarization maps that will be soon published in \cite{planck2018}. The colour correction has been accounted for according to \cite{planckhfispectral}.}
    \label{tab:planck_results}
 \end{table*}

\subsection{Polarization degree and angle estimates}
In order to compare our results with low angular resolution CMB experiments, we present in Table~\ref{tab:crab_results} the integrated Stokes $I$, $Q$, and $U$, and the associated polarization intensity I$_\textrm{pol}$, and polarization degree $p$ and angle $\psi$. A colour correction factor of 1.05 has been accounted for. All the listed values are estimated using aperture photometry inside apertures of 5$^\prime$, 7$^\prime$, and 9$^\prime$ from the centre of the map. We limit the analysis to the maximum of the map size avoiding the edges and considering that all the flux is concentrated in 7.6' of aperture (see Fig.~\ref{crab_integrated_flux}). The uncertainties account for the calibration error of 10\% discussed above. 
The polarization efficiency factor estimated across the \NIKA\ 2.05 mm spectral band and reported in \cite{ritacco2017} is $\rho_{pol}$ = 0.9941 $\pm$ 0.0002. This very small efficiency loss of 0.6 \% has a negligible impact on the estimation of the polarization fluxes and the calibration error itself. 
The polarization angle is  presented here in equatorial coordinates and Galactic coordinates to ease the comparison with CMB experiments.
The polarization angle uncertainty accounts for 2.3$^{\circ}$ systematic uncertainties, while the statistical uncertainties also accounts  for Monte Carlo simulations of the noise in $Q$ and $U$ and the difference between two sets of jack-knife noise maps (seven maps each).

Figure~\ref{crab_p_angle_comparison} shows the polarization fraction (top) and polarization angle (bottom) of the \object{Crab} nebula as a function of the frequency as measured by five different instruments: 
\WMAP\ \citep{2011ApJS..192...19W}, XPOL \citep{aumont2010}, POLKA \citep{2014PASP..126.1027W}, \Planck\ \citep{2015arXiv150702058P}, and \NIKA\ (this paper). 
We note that the \WMAP\ satellite has FWHMs: 0.93$^{\circ}$, 0.68$^{\circ}$, 0.53$^{\circ}$, 0.35$^{\circ}$, and \textless 0.23$^{\circ}$ at 22 GHz, 30 GHz, 40 GHz, 60 GHz, and 90 GHz, respectively. XPOL and POLKA have FWHMs of 27$^{\prime\prime}$ and 20$^{\prime\prime}$, respectively.
The \Planck\ satellite FWHMs are 33$^{\prime}$, 24$^{\prime}$, 14$^{\prime}$, 10$^{\prime}$, 7.1$^{\prime}$, 5.5$^{\prime}$, and 5$^{\prime}$ at 30, 44, 70, 100, 143, 217, and 353 GHz, respectively. Furthermore, after discussions with the \Planck\ team we  reanalysed the \Planck\ HFI data using the polarization maps that will be soon published in \cite{planck2018}. We have performed aperture photometry directly in the Healpix maps. The results are given in Table~\ref{tab:planck_results}.
The \NIKA\ and POLKA values in Fig.~\ref{crab_p_angle_comparison} have been estimated by aperture photometry up to 9$^{\prime}$. The XPOL value considers 10$^{\prime}$ \citep{aumont2010} and the other experiments their native FWHM.

Using all the data sets and accounting for systematics as indicated in \citet{2011ApJS..192...19W,thum2008} and \citet{rosset2010} for \WMAP, XPOL and \Planck\ respectively,  we compute the weighted-average of the polarization angle $\psi$ = $-87.7^{\circ}\pm 0.3^{\circ}$.  
All the observations shown on the bottom panel of
Fig.~\ref{crab_p_angle_comparison} agree with this value within 1$\sigma$,  except for POLKA.
In addition we find  good agreement between  \NIKA\ and POLKA and we also find that \NIKA\ is consistent within 1$\sigma$ with the \Planck\ value at 143 GHz. The \NIKA\ result differs from the average value by $\sim$3$^{\circ}$.  This result could in principle be explained by an error in the calibration of the polarization angle.
In \cite{ritacco2017} a calibration angle error of 3$^{\circ}$ was also considered to explain the difference  in the polarization angle of calibration sources between \NIKA\ and the other experiments. However, the results presented in \cite{ritacco2017} were consistent within the error bars with the other experiments
and did not allow us to accurately estimate this possible shift in the angle. 
As  \NIKA\ is no longer in use we have no means of measuring this calibration error.
  
Using all the available data sets except the POLKA results, we have computed the weighted average degree of polarization and uncertanties on it.
We find 6.95$\pm$0.03$\%$, as shown by the solid line and dashed lines in the top panel of Fig.~\ref{crab_p_angle_comparison}. We observe that most of the results between 20 and 353 GHz are consistent with this value at the 1$\sigma$ level, except for \Planck\ at 70 GHz, XPOL, and POLKA,  which show a significantly larger degree of polarization.

\begin{figure}
  \centering
          { \includegraphics[width=1\linewidth,keepaspectratio]{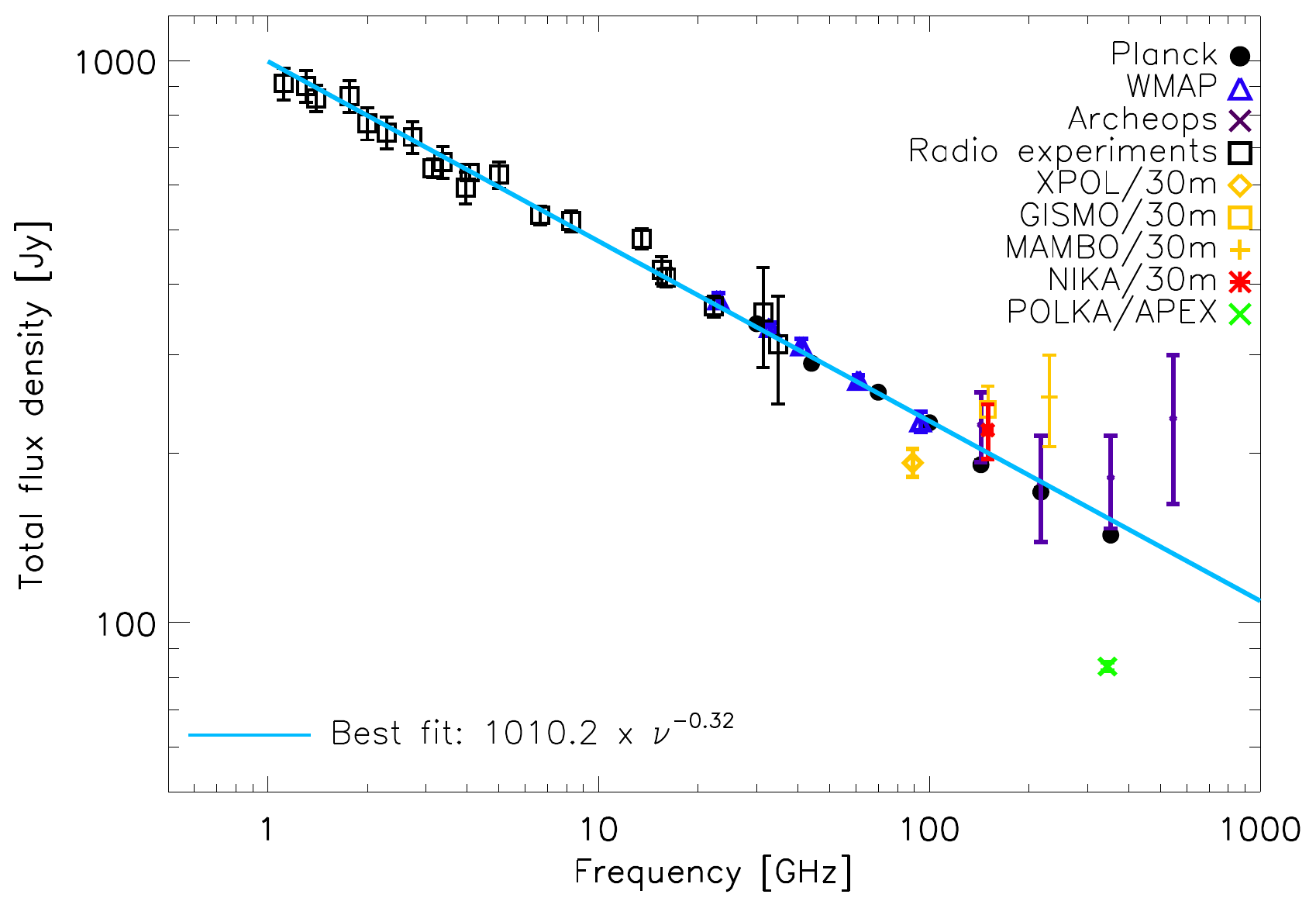}}
           \caption{\object{Crab} nebula total power SED as obtained from \Planck\ \citep{2015arXiv150702058P} for the LFI instrument, \Planck\ HFI data reanalysed from the maps that will be soon published in \cite{planck2018}, \WMAP\ \citep{2011ApJS..192...19W}, \Archeops\ \citep{macias2007archeops}, radio experiments \citep{dmitrenko1970absolute, 1971IzVUZ..14..157V}, XPOL/30m \citep{aumont2010}, \NIKA/30m (this paper), MAMBO/30m \citep{2002A&A...386.1044B}, POLKA/APEX \citep{2014PASP..126.1027W}, and GISMO/30m \citep{2011ApJ...734...54A} data. \NIKA\ and POLKA values are estimated over the entire extent of the source. The best-fit single power law model obtained by the analysis in this paper is shown in cyan. The best-fit models and the data both account for the \object{Crab} nebula fading with time, using 2018 as year of reference. The POLKA data flux loss ($\sim$40\%) is compatible with the losses expected due to the spatial filtering of total intensity in the LABOCA data reduction \citep{2011A&A...527A.145B}.}
\label{crab_SED}                
  \end{figure}

For XPOL the discrepancy can
probably be explained by the lower sensitivity of the single channel XPOL experiment to the lower-than-average polarization of the outer parts of the
nebula.
POLKA shows a very high polarization degree due to the $\sim$40\% flux loss observed in Stokes $I$ (see Fig.~\ref{crab_SED}). This is compatible with the losses expected due to the spatial filtering of total intensity in LABOCA data reduction in this range of angular scales \citep{2011A&A...527A.145B}.
In the case of \Planck\ HFI, we only find significant discrepancies for the 143 GHz data that remain unexplained to date.

The relatively constant behaviour of the polarization degree and angle over a wide frequency range suggests that the polarization emission is driven by the same physical process. We therefore expect a well-defined SED for the \object{Crab} nebula intensity and polarization emission (see next section).

\section{Characterization of the Crab spectral energy distribution in intensity and polarization}\label{sec:Polarization intensity Spectral Energy Density (SED)}
\subsection{Intensity}
The total flux density of the \object{Crab} nebula at radio and millimetre
wavelengths (from 1 to 500 GHz) is mainly expected to be due to synchrotron emission and can be
well described by a single power law of the form
\begin{equation}
I_{\nu} = A(\nu / 1 {\bf GHz})^{\beta}
\label{eq:sync1}
\end{equation}
with spectral index $\beta$ = --0.296$\pm$0.006 \citep{baars1977absolute,macias2010}. Further, the emission of the \object{Crab} nebula is fading with time at a rate of $\alpha$ = --0.167$\pm$0.015 \% yr$^{-1}$ \citep{aller1985decrease}. 
These results suggest a low frequency emission produced by particles accelerated by the same magnetic field. \cite{macias2010} have also shown  that there is no evidence for an extra synchrotron component or for thermal dust emission at these frequencies. The direction and degree of the polarization is therefore expected to be constant across the frequency range 30--300 GHz.

Figure~\ref{crab_SED} shows the total flux density of the \object{Crab} nebula as a function of frequency. The fluxes in the radio domain were taken from \cite{dmitrenko1970absolute} and \cite{1971IzVUZ..14..157V}. We also show microwave and millimetre wavelength fluxes from \Archeops\ \citep{macias2007archeops}, \Planck\ \citep{2015arXiv150702058P}, \WMAP\ \citep{2011ApJS..192...19W}, XPOL \citep{aumont2010}, MAMBO/30m \citep{2002A&A...386.1044B}, POLKA \citep{2014PASP..126.1027W}, and GISMO/30m \citep{2011ApJ...734...54A}. 
The HFI \Planck\ fluxes were specifically estimated for this paper using the new
2018 Planck HFI intensity and polarization maps, which will be made publicly available before
the end of the year. We note that in these new maps the treatment of the polarization systematics has been significantly improved with respect to those in~\cite{2015arXiv150702058P}. Furthermore, the Planck HFI \object{Crab} nebula fluxes in~\cite{2015arXiv150702058P} were computed assuming a point source, which is not adapted
for an extended source.
The measured \NIKA\ total flux density at 150 GHz is shown in red.  We note that in the plot both the best-fit model and the data represented are corrected for the fading of the source.

Assuming the single power law model in Eq.~\ref{eq:sync1} and
by $\chi^2$ minimization we obtain
$$\setlength\arraycolsep{0.1em}
 \begin{array}{rclcl}
  \textrm{A}&=& 1010.2\pm3.8 & & \textrm{Jy}; \quad \quad  \textrm{$\beta$} = - 0.323\pm0.001\\
 \end{array}
$$
The best-fit model is shown in Fig.~\ref{crab_SED} in cyan.
The \NIKA\ data are consistent with this model at the 1$\sigma$ level.
The estimated spectral index $\beta$ is slightly different from the previous results provided by \cite{macias2010}. This 
is probably due to the addition of new \Planck\ and  WMAP data.

As already discussed above, XPOL total power emission is low with respect to expectations.
The POLKA value is found to be  lower than the \Planck\ result at the same frequency; this is mainly explained by the spatial filtering of LABOCA data reduction, as already discussed in the previous section.

\subsection{Polarization}
\begin{figure}
  \centering
             { \includegraphics[width=1\linewidth,keepaspectratio]{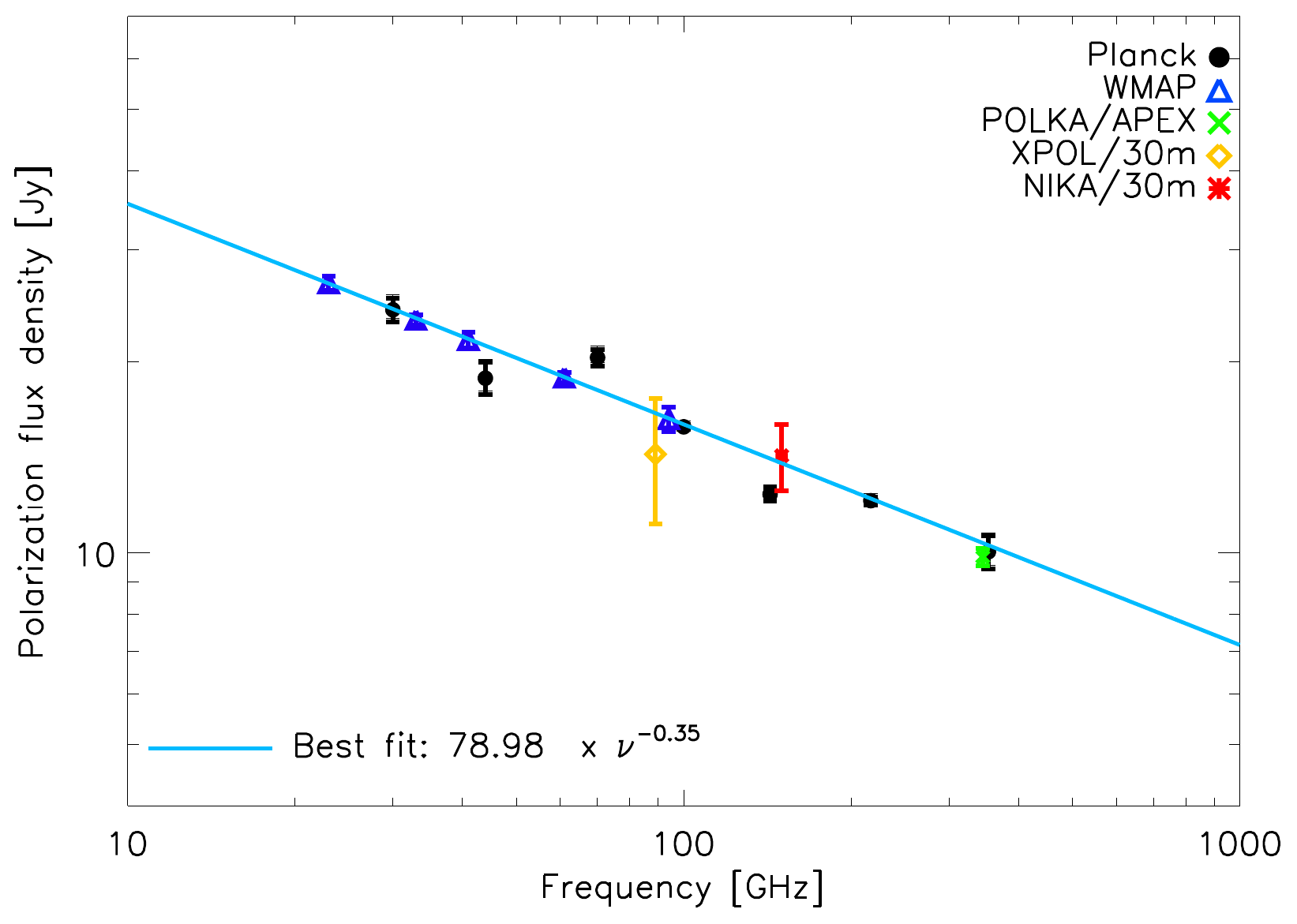}}
           \caption{\object{Crab} nebula polarization flux SED as obtained from \Planck\ LFI \citep{2015arXiv150702058P}, \Planck\ HFI data reanalysed from the maps soon published in \cite{planck2018}, \WMAP\ \citep{2011ApJS..192...19W}, XPOL \citep{aumont2010}, POLKA \citep{2014PASP..126.1027W}, and \NIKA\ (this paper) data. \NIKA\ and POLKA values are estimated over the entire extent of the source. The best-fit models and the data both account for the \object{Crab} nebula fading with time, using 2018 as year of reference. We also show the single power law best-fit model in cyan.}
\label{crab_SED_ipol}           
  \end{figure}

Though the total power emission of the \object{Crab} nebula has been monitored over decades across a wide range of frequencies, the amount of polarization data is still poor.
Recent results from the \Planck\ LFI instrument \citep{2015arXiv150702058P}, the new \Planck\ HFI maps presented above \citep{planck2018}, \WMAP\ \citep{2011ApJS..192...19W},
XPOL \citep{aumont2010}, and POLKA \citep{2014PASP..126.1027W} data, together with the \NIKA\ results allow us to trace the SED of the polarized emission, I$_{\textrm{pol}}$, of the \object{Crab} nebula as shown in Fig.~\ref{crab_SED_ipol}.  
We note that the uncertainties for the \NIKA\ polarization intensity also include  absolute calibration errors
and systematics as discussed in the previous sections.  
Assuming a single power law synchrotron emission (see Eq.~\ref{eq:sync1}) for the polarization emission of the \object{Crab} nebula and a using $\chi^2$ fitting procedure we find
$$
\setlength\arraycolsep{0.1em}
 \begin{array}{rclcl}
  \textrm{A$_{pol}$}&=& 78.98\pm7.82 & & \textrm{Jy}; \quad \textrm{$\beta_{pol}$} = - 0.347\pm0.026\\
 \end{array}
 $$
 
We observe that the \NIKA, XPOL, and POLKA results are consistent with the best-fit model at the 1$\sigma$ level.
We have also estimated the spectral index of the \object{Crab} nebula polarization emission at high frequency using the map obtained by SCUPOL \citep{scubapol} at 352 GHz (850
$\mu$m) and the \NIKA\ map. Considering only the region observed by SCUPOL ($\sim$1.5$^{\prime}$) we
obtain $\beta_{pol}^{\small NIKA/SCUPOL} = -0.33 \pm 0.01$.
This result is in good agreement with the best-fit model spectral index presented above.

The polarization spectral index is consistent with the total power index confirming that the synchrotron radiation is the fundamental mechanism that drives the polarization emission of the \object{Crab} nebula.

 \noindent

\section{Conclusions}\label{sec:conclusions}
The \object{Crab} nebula is considered  a celestial standard calibrator for CMB experiments in
terms of polarization degree and angle. An absolute calibration is
particularly important for reliable measurements of the CMB polarization B-modes,
which are a window towards the physics of the early Universe.

We have reported in this paper the first high angular resolution polarization observations of the \object{Crab} nebula at 150 GHz, which were obtained with the \NIKA\ camera. These observations have
allowed us to map the spatial distribution of the \object{Crab} nebula polarization
fraction and angle.  

Using the \NIKA\ data, in addition to all the available polarization data to date, we conclude that the
polarization angle of the \object{Crab} nebula is consistent with being constant with
frequency, from 20 GHz to 353 GHz, at arcmin scales with a value of
$-87.7^{\circ}\pm0.3$ in Galactic coordinates. High resolution observations provided by \NIKA\ at 150 GHz and POLKA at 345 GHz show a polarization angle that is lower than the average value by $\sim 3^{\circ}$ and $\sim 5^{\circ}$, respectively.
Though the uncertainties on these values are high because of the systematic errors, this discrepancy highlights the need of further high angular resolution polarization observations in this frequency range.
In addition, we find a strong case for a constant polarization degree of $p$ = 6.95 $\pm$ 0.03\%. 

Moreover, we have characterized the intensity and polarization SED of the \object{Crab} nebula. In both total power and polarization, we find that the data are overall consistent with a single power law spectrum, as expected from synchrotron emission from a single population of relativistic electrons.The \object{Crab} nebula presents a polarization spectral index $\beta_{pol}=-0.347 \pm 0.026$ that is consistent with the intensity spectral index $\beta=-0.324 \pm 0.001$. 
However, we find some discrepancies between the  data sets which will require further millimetre measurements. Among future polarization experiments, \NIKAd\ \citep{calvo2016,2017arXiv170700908A}, will provide high sensitive polarization observations of the \object{Crab} nebula adding a 260~GHz map at 11$^{\prime\prime}$ resolution.

\bibliography{bib_crab}

\vspace{0.2cm}
 \begin{acknowledgements}
We would like to thank the IRAM staff for their support during the \NIKA\ campaigns. 
The NIKA dilution cryostat has been designed and built at the Institut N\'eel. 
In particular, we acknowledge the crucial contribution of the Cryogenics Group, and 
in particular Gregory Garde, Henri Rodenas, Jean Paul Leggeri, and Philippe Camus. 
This work has been partially funded by the Foundation Nanoscience Grenoble, the LabEx FOCUS ANR-11-LABX-0013, and 
the ANR under the contracts ``MKIDS'', ``NIKA'', and ANR-15-CE31-0017. 
This work has benefited from the support of the European Research Council Advanced Grant ORISTARS 
under the European Union's Seventh Framework Programme (Grant Agreement no. 291294).
We thank the Planck Collaboration for allowing us to use the 2018 Planck maps in advance of public release to obtain integrated flux densities in intensity and polarization in the Crab nebula. We acknowledge funding from the ENIGMASS French LabEx (R.A. and F.R.) and
the CNES post-doctoral fellowship program (R.A.). We acknowledge  support from the Spanish Ministerio de Econom\'ia and Competitividad (MINECO) through grant number AYA2015-66211-C2-2 (R.A.), the CNES doctoral fellowship program (A. R.), and
the FOCUS French LabEx doctoral fellowship programme (A.R.).
A.M. has received funding from the European Research Council (ERC) under the European Union’s Horizon 2020 research and innovation programme (Magnetic YSOs, grant agreement no. 679937).
\end{acknowledgements}

\end{document}